\documentclass[aps,prb,11pt,showpacs]{revtex4-1}
\usepackage{mathrsfs}
\usepackage{amsmath}
\usepackage{amsfonts}
\usepackage{amssymb}
\usepackage{amsthm}
\usepackage{graphicx}
\usepackage{color}
\usepackage{hyperref}
\usepackage{bm}
\usepackage[caption=false]{subfig}
\usepackage{verbatim}

\begin{document}
\title{Magnon-mediated spin current noise in ferromagnet$|$non-magnetic conductor hybrids}
\author{Akashdeep Kamra}
\email{akashdeep.kamra@uni-konstanz.de} 
\author{Wolfgang Belzig}
\email{wolfgang.belzig@uni-konstanz.de}
\affiliation{Fachbereich Physik, Universit{\"a}t Konstanz, D-78457 Konstanz, Germany}

\begin{abstract}
The quantum excitations of the collective magnetization dynamics in a ferromagnet (F) - magnons - enable spin transport without an associated charge current. This pure spin current can be transferred to electrons in an adjacent non-magnetic conductor (N). We evaluate the finite temperature noise of the magnon-mediated spin current injected into N by an adjacent F driven by a coherent microwave field. We find that the dipolar interaction leads to squeezing of the magnon modes giving them wavevector dependent non-integral spin, which directly manifests itself in the shot noise. For temperatures higher than the magnon gap, the thermal noise is dominated by large wavevector magnons which exhibit negligible squeezing. The noise spectrum is white up to the frequency corresponding to the maximum of the temperature or the magnon gap. At larger frequencies, the noise is dominated by vacuum fluctuations. The shot noise is found to be much larger than its thermal counterpart over a broad temperature range, making the former easier to be measured experimentally. 
\end{abstract} 




\maketitle

\section{Introduction}\label{sec:intro}
Interest in magnetic nanostructures has been motivated, in part, by their numerous applications in the electronics industry. Starting with metallic magnets, there has been a recent upsurge of interest in magnetic insulators because of their low Gilbert damping. The latter is understood as due to the absence of conduction electrons which typically constitute the dominant scattering channel for {\em magnons} - the elementary excitations representing collective magnetization dynamics. Furthermore, magnons carry spin without an associated charge, which can conveniently be transferred to the electronic degrees of freedom in a ferromagnet (F)$|$ non-magnetic conductor (N) bilayer~\cite{Tserkovnyak2002,Weiler2013}. New transport paradigms based on magnons, instead of electrons, have emerged~\cite{Kruglyak2010,Shindou2013}. While the two kinds of quasi-particles share similarities due to their typically parabolic dispersion relations, the bosonic nature of the magnons offers new unique possibilities~\cite{Demokritov2006}.

A magnet can exchange spin current only in directions orthogonal to its magnetic moment~\cite{Brataas2012}. However, at finite temperatures, the latter fluctuates around its equilibrium orientation and thus, on an average, allows a ``longitudinal''spin current absorption and emission. When the magnet is insulating, this spin transfer can be ascribed entirely to magnons. Even for metallic magnets, magnonic contribution may dominate over its electronic counterpart~\cite{Uchida2010,Xiao2010,Adachi2013}. With an increasing emphasis on magnonic~\cite{Kruglyak2010} and caloric~\cite{Bauer2012} phenomena, finite temperature effects cannot be disregarded and have taken the center stage in several investigations~\cite{Bender2015,Xiao2015}. 

Non-zero temperatures, on the other hand, make it necessary to consider fluctuations, often referred to as noise, in physical quantities. While the magnetization fluctuations are well studied~\cite{Brown1963,Safonov2002,Foros2005,Rossi2005}, pure spin current noise has received attention only recently~\cite{Belzig2004,Kamra2014}. Non-equilibrium spin accumulation has been shown to result in charge current shot noise~\cite{Arakawa2015}. The (inverse) spin Hall effect (SHE) mediated spin-charge current conversion offers a convenient method to measure spin currents~\cite{Hirsch1999}. This has been exploited in the observation of the thermal pure spin current noise in a yttrium iron garnet (YIG)$|$platinum (Pt) heterostructure~\cite{Kamra2014}. However, owing to the fluctuation-dissipation theorem~\cite{Callen1951}, information obtained via thermal noise is also accessible via the typically easier to measure linear response of the system. Non-equilibrium noise, on the other hand, delineates microscopic dynamics not accessible via the observable average~\cite{Blanter2000,Beenakker2003,Nazarov2003}. For example, charge current shot noise has been instrumental in, among several phenomena, ascertaining unconventional quanta of charge transport in different exotic phases of interacting electronic systems~\cite{Jain1989,Kane1994,Reznikov1999,Jehl2000}. In a similar fashion, spin current shot noise can be exploited to probe the quantum of spin transport. We have recently demonstrated that the zero-temperature shot noise of spin current across an F$|$N interface indicates spin transport in non-integral quanta~\cite{Kamra2016}. 

In the present work, we evaluate the finite temperature noise of the magnon-mediated spin current traversing the F$|$N interface, when F is driven by a microwave magnetic field. The resulting total noise is composed of the shot noise, stemming from the discrete nature of the microwave driven spin transfer, and the thermal noise caused by the dynamic spin exchange between the equilibrium magnons in F and electrons in N. A key finding is that, in contrast to typical electronic systems~\cite{Blanter2000}, the spin current shot noise in our system increases linearly with temperature and dominates the total noise over a broad experimental parameter space. This is attributed to the large number of magnonic excitations created by the microwave drive in comparison with the relatively small number of thermal excitations in F, a feature which is unique to a non-conserved boson gas.

Owing to the dipolar interaction, the eigenmodes of F are squeezed-magnons (s-magnons) which possess, wavevector and applied magnetic field dependent, non-integral spins. The squeezing is maximum for the low energy magnons while it decreases with increasing relative contribution of the exchange for high wavenumbers. Thus, the dipolar interaction significantly influences the shot noise, which is attributed to the non-equilibrium zero wavevector s-magnons possessing a non-integral spin $\hbar^* = \hbar (1 + \delta)$. On the other hand, barring very low temperatures, dipolar interaction can be disregarded in evaluating the thermal noise, which has contributions from a broad region of the wavevector space. Thus, in addition to exact numerical evaluation, we obtain analytical results for the thermal, including the vacuum, noise in various limiting cases finding good agreement with numerics. Vanishing for frequencies below the magnon gap, the vacuum noise dominates the total noise power at large frequencies. The thermal and shot contributions to the noise are white up to about the frequency corresponding to the larger between the temperature and the magnon gap, increasing with frequency thereafter.

The paper is organized as follows. Section \ref{sec:system} describes the system under investigation and the theoretical method employed to evaluate the physical quantities of interest. It is further divided into subsections with a detailed derivation of the Hamiltonian in subsection \ref{ssec:hamiltonian}, discussion of the dynamical equations of motion in subsection \ref{ssec:eom}, and a derivation of the general expression for the spin current noise in subsection \ref{ssec:neval}. The final expressions obtained for the shot (subsection \ref{ssec:neqres}) and the thermal (subsection \ref{ssec:eqres}) noises are reported in section \ref{sec:results}. We discuss the relevance of our results putting them in a broader context in section \ref{sec:discussion}. Finally, we conclude by summarizing our work in section \ref{sec:summary}.

\section{System and theoretical framework}\label{sec:system}

\begin{figure}[htb]
\begin{center}
\includegraphics[width=65mm]{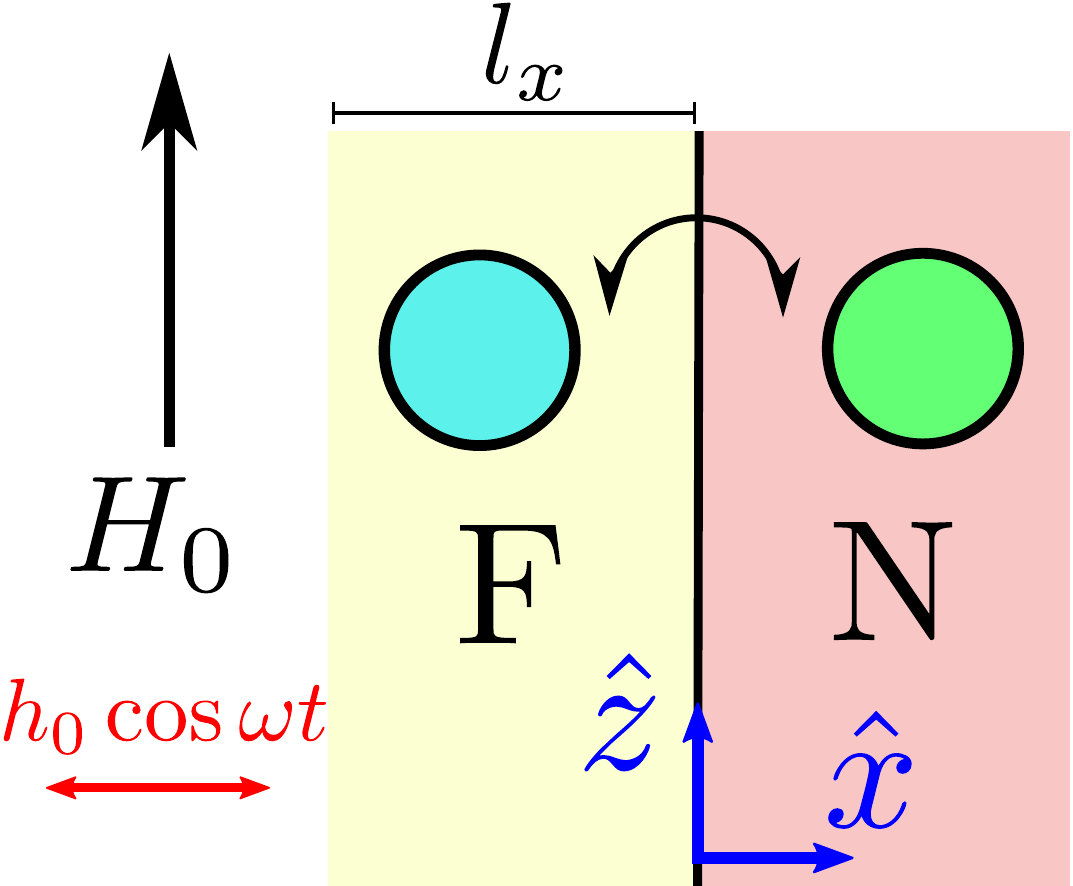}
\caption{System schematic. An applied static magnetic field ($H_0~\hat{\pmb{z}}$) saturates the magnetization of the ferromagnet (F) along the z-direction. An oscillating magnetic field ($h_0 \cos \omega t ~\hat{\pmb{x}}$) creates non-equilibrium, in addition to thermal, magnonic excitations in F, which annihilate at latter's interface with a non-magnetic conductor (N) creating new excitations and transferring spin current.}
\label{fig:bilayer}
\end{center}
\end{figure}

We consider a F$|$N bilayer (Figure \ref{fig:bilayer}) subjected to a static magnetic field $H_0~\hat{\pmb{z}}$ which saturates the equilibrium magnetization of F along the z-direction. At finite temperatures, the F magnetic moment fluctuates about its equilibrium orientation which can be represented by thermal magnonic excitations. The latter dynamically exchange spin with the electrons in N giving rise to a fluctuating spin current across the interface. A microwave magnetic field $h_0 \cos \omega t ~\hat{\pmb{x}}$ additionally creates non-equilibrium magnetization dynamics resulting in a net spin current flow into N and an associated shot noise. For metallic F, the additional contribution to the spin current noise due to spin exchange between F and N conduction electrons is {\em not} considered here.  

Our methodology entails obtaining the system Hamiltonian and the spin current operator in terms of the creation and annihilation operators of the magnonic and electronic eigenmodes in F and N, respectively. Thereafter, Heisenberg equations of motion are employed to evaluate the microwave field driven coherent magnetization dynamics as well as the time evolution and noise of the spin current traversing the F$|$N interface. 

\subsection{Hamiltonian}\label{ssec:hamiltonian}
The total Hamiltonian comprises of the terms due to the magnetic degrees of freedom in F, electrons in N, interaction between F magnetization and N electrons, and the driving of the F magnetization by the coherent microwave field:
\begin{align}\label{htot}
\tilde{\mathcal{H}} & = \tilde{\mathcal{H}}_{\mathrm{F}} + \tilde{\mathcal{H}}_{\mathrm{N}} + \tilde{\mathcal{H}}_{\mathrm{int}} + \tilde{\mathcal{H}}_{\mathrm{drive}},
\end{align}
where {\em we use tilde to denote operators}. For simplicity, we do not explicitly consider the non-linear terms in $\tilde{\mathcal{H}}_{\mathrm{F}}$ and $\tilde{\mathcal{H}}_{\mathrm{N}}$ that are responsible for dissipation and equilibration in the two subsystems.

\subsubsection{Magnetic contribution}\label{sssec:magcont}

We employ the `macroscopic magnon theory'~\cite{Kittel1963} in describing the collective magnetization eigenmodes and their dynamics in F. This formalism allows a quantum treatment based on the general phenomenological theories of magnetism without reference to a definite microscopic model. Hence, it affords a wide applicability, within the low wavenumber limit, while yielding results identical to those obtained from the microscopic model~\cite{Akhiezer1968}, when the latter constitutes a valid description of the material system under consideration. 

We first write the classical magnetic free energy $\mathcal{H}_{\mathrm{F}}$ which, in turn, is constituted by Zeeman , anisotropy, exchange and dipolar interaction energy densities:
\begin{align}\label{mhamilclass}
\mathcal{H}_{\mathrm{F}} & =  \int_{V_{\mathrm{F}}} d^3 r \left( H_{\mathrm{Z}} + H_{\mathrm{aniso}} + H_{\mathrm{ex}} + H_{\mathrm{dip}}  \right) , 
\end{align}   
where $V_{\mathrm{F}}$ is the volume of F. Expanding the free energy densities about the equilibrium configuration $\pmb{M} = M_s \hat{\pmb{z}}$, with $\pmb{M}$ and $M_s$ respectively the magnetization and saturation magnetization, retaining terms up to the second order in the field variables $M_{x,y}~(\ll M_z \approx M_s)$~\cite{Kittel1949,Kamra2015}:
\begin{align}
H_{\mathrm{Z}} + H_{\mathrm{aniso}} & = \frac{\omega_{\mathrm{za}}}{2 |\gamma| M_s} \left( M_x^2 + M_y^2 \right),
\end{align}
with $\omega_{\mathrm{za}} = |\gamma|[\mu_0 H_0 + 2(K_1 + K_u)/M_s]$, where $\gamma$ is the typically negative gyromagnetic ratio, $\mu_0$ is the permeability of free space, and $K_{u} (>0)$ and $K_{1} (>0)$, respectively, represent contributions from easy axes uniaxial and cubic magnetocrystalline anisotropies. The exchange energy density for a cubic crystal is parameterized in terms of the exchange constant $A$~\cite{Kittel1949}:
\begin{align}
H_{\mathrm{ex}} & = \frac{A}{M_s^2} \left[  \left( \pmb{\nabla} M_x \right)^2  + \left( \pmb{\nabla} M_y \right)^2     \right] . 
\end{align}
The dipolar interaction can be treated within a mean field approximation via the so-called demagnetization field $\pmb{H}_m$ generated by the magnetization:
\begin{align}
 H_{\mathrm{dip}} & = - \frac{1}{2} \mu_0 \pmb{H}_m \cdot \pmb{M}.
\end{align}
The magnetization and the demagnetization field are split into spatially uniform and non-uniform contributions $\pmb{H}_m   = \pmb{H}_{u} + \pmb{H}_{nu}$ and $\pmb{M}   = \pmb{M}_{u} + \pmb{M}_{nu}$ thereby affording the following relation between the uniform components~\cite{Akhiezer1968,Kittel1963}:
\begin{align}
\pmb{H}_{u} = - N_x M_{ux}~ \hat{\pmb{x}} -  N_y M_{uy} ~\hat{\pmb{y}} - N_z M_{uz} ~\hat{\pmb{z}},
\end{align} 
where $N_{x,y,z}$ are the eigenvalues of the demagnetization tensor which is diagonal in the chosen coordinate system. Within the magnetostatic approximation~\footnote{Strictly speaking, the magnetostatic approximation is not valid for a certain narrow range of low k excitations~\cite{Akhiezer1968}. However, as we see in the final results, the thermal noise has contribution from excitations in a wide k range making the error due to an imprecise treatment of a fraction of this range negligible.}, the non-uniform components obey the equations~\cite{Akhiezer1968,Kittel1963}:
\begin{align}
\pmb{\nabla} \times \pmb{H}_{nu} & = 0, \\
\pmb{\nabla} \cdot \left( \pmb{H}_{nu} + \pmb{M}_{nu} \right) & = 0. \label{divdemagfield}
\end{align}
Employing the equations above and Fourier representation, the dipolar interaction energy can be written as a sum over the k space, as will be presented below. 

The quantization of the classical magnetic Hamiltonian is achieved by defining the magnetization operator $\tilde{\pmb{M}} = - |\gamma| \tilde{\pmb{S}}_{\mathrm{F}}$ in terms of the spin density operator in F: $\tilde{\pmb{S}}_{\mathrm{F}}$, where we have assumed a negative gyromagnetic ratio $\gamma$. Employing the general commutation relations between the components of angular momentum, we obtain:
\begin{align}
\left[ \tilde{M}_{+} (\pmb{r}), \tilde{M}_{-} (\pmb{r}^\prime)  \right] & = 2 |\gamma| \hbar \tilde{M}_z (\pmb{r}) ~ \delta (\pmb{r} - \pmb{r}^\prime),
\end{align}
with $\tilde{M}_{\pm} = \tilde{M}_x \pm i (\gamma / |\gamma|) \tilde{M}_y$~\footnote{The $\gamma / |\gamma|$ factor, which is often omitted assuming positive $\gamma$, is essential for a valid transformation consistent with angular momentum conservation.}. These commutation relations are satisfied by the Holstein-Primakoff transformations~\cite{Holstein1940,Kittel1963} relating the magnetization operator to the bosonic creation and annihilation operators $\tilde{a}^\dagger(\pmb{r}), \tilde{a}(\pmb{r})$:
\begin{align}
\tilde{M}_{+} & = \sqrt{2 |\gamma| \hbar M_s} \left( 1 - \frac{|\gamma| \hbar}{2 M_s} \tilde{a}^\dagger \tilde{a} \right)^\frac{1}{2} \tilde{a} \ \approx \ \sqrt{2 |\gamma| \hbar M_s} \ \tilde{a} , \\
\tilde{M}_{-} & = \sqrt{2 |\gamma| \hbar M_s} \tilde{a}^\dagger \left( 1 - \frac{|\gamma| \hbar}{2 M_s} \tilde{a}^\dagger \tilde{a} \right)^\frac{1}{2} \ \approx \ \sqrt{2 |\gamma| \hbar M_s} \ \tilde{a}^\dagger , \\
\tilde{M}_z & = M_s - |\gamma| \hbar \tilde{a}^\dagger \tilde{a}.
\end{align}
Here, $\tilde{a}^\dagger(\pmb{r})$ flips the spin at position $\pmb{r}$ thereby creating a localized magnonic excitation, and is related to the normal magnon operators $\tilde{b}_{\pmb{q}}^\dagger$ via $\tilde{a}^\dagger(\pmb{r}) = \sum_{\pmb{q}} \phi_q^* (\pmb{r}) \tilde{b}^\dagger_{\pmb{q}}$ with plane wave eigenstates $\phi_{\pmb{q}} (\pmb{r}) = (1/\sqrt{V_{\mathrm{F}}}) \exp (i \pmb{q} \cdot \pmb{r})$. Thus, up to the first order in operators, the components of the magnetization operator can be written in the Fourier space:
\begin{align}
\tilde{M}_x & =  \sum_{\pmb{q}} \sqrt{\frac{|\gamma| \hbar M_s}{2 V_{\mathrm{F}}}} \left( \tilde{b}_{-\pmb{q}}^\dagger  + \tilde{b}_{\pmb{q}} \right) e^{i \pmb{q} \cdot \pmb{r}}    , \label{mxquant} \\
\tilde{M}_y & =  \sum_{\pmb{q}} \frac{1}{i} \sqrt{\frac{|\gamma| \hbar M_s}{2 V_{\mathrm{F}}}} \left( \tilde{b}_{-\pmb{q}}^\dagger  - \tilde{b}_{\pmb{q}} \right) e^{i \pmb{q} \cdot \pmb{r}} . \label{myquant}
\end{align}
Employing the above two equations (\ref{mxquant}) and (\ref{myquant}) into equations (\ref{mhamilclass}) to (\ref{divdemagfield}) and disregarding the zero-point energy, we obtain the magnetic Hamiltonian bilinear in the k-space magnon operators:
\begin{align}
\tilde{\mathcal{H}}_{\mathrm{F}} & = \sum_{\pmb{q}} \left( A_{\pmb{q}} ~ \tilde{b}^\dagger_{\pmb{q}} \tilde{b}_{\pmb{q}} +  B_{\pmb{q}}^* ~ \tilde{b}_{\pmb{q}}^\dagger \tilde{b}_{-\pmb{q}}^\dagger + B_{\pmb{q}} ~ \tilde{b}_{\pmb{q}} \tilde{b}_{-\pmb{q}} \right),
\end{align} 
where
\begin{align}
A_{\pmb{q}} \ = \ A_{-\pmb{q}} & =  \hbar \left(  \omega_{\mathrm{za}} - \omega_s N_z + D q^2 + \frac{\omega_s}{2} (N_x + N_y) \delta_{\pmb{q},\pmb{0}} + (1 - \delta_{\pmb{q},\pmb{0}}) \frac{\omega_s}{2} \sin^2 \theta_{\pmb{q}}  \right) , \\
B_{\pmb{q}} \ = \ B_{-\pmb{q}} & =  \hbar \left( \frac{\omega_s}{4} N_{xy} \delta_{\pmb{q},\pmb{0}} + (1 - \delta_{\pmb{q},\pmb{0}}) \frac{\omega_s}{4} \sin^2 \theta_{\pmb{q}} ~ e^{i 2 \phi_{\pmb{q}}} \right) .
\end{align}
Here, $D = 2 A |\gamma|/M_s$, $\omega_s = |\gamma| \mu_0 M_s$, $N_{xy} = N_x - N_y$, $\theta_{\pmb{q}}$ and $\phi_{\pmb{q}}$ are respectively the polar and azimuthal angles of the wavevector $\pmb{q}$. The magnetic Hamiltonian thus obtained may be brought to a diagonal form by the Bogoliubov transformations~\cite{Holstein1940,Akhiezer1968} to new bosonic quasi-particles corresponding to the annihilation operators $\tilde{\beta}_{\pmb{q}} = u_{\pmb{q}} \tilde{b}_{\pmb{q}} - v_{\pmb{q}}^* \tilde{b}^\dagger_{- \pmb{q}}$:
\begin{align}\label{hf}
\tilde{\mathcal{H}}_{\mathrm{F}} & = \sum_{\pmb{q}}  \hbar \omega_{\pmb{q}} \tilde{\beta}^\dagger_{\pmb{q}} \tilde{\beta}_{\pmb{q}},
\end{align}
with the transformation parameters $\hbar \omega_{\pmb{q}} = \sqrt{A_{\pmb{q}}^2 - 4 |B_{\pmb{q}}|^2 }$, and
\begin{align}
v_{\pmb{q}}  = - \frac{2 B_{\pmb{q}}}{(A_{\pmb{q}} + \hbar \omega_{\pmb{q}})} \ u_{\pmb{q}}  & = - e^{i \Theta_{\pmb{q}}} \frac{2 B_{\pmb{q}}}{\sqrt{(A_{\pmb{q}} + \hbar \omega_{\pmb{q}})^2 - 4 |B_{\pmb{q}}|^2}}.
\end{align}
$\Theta_{\pmb{q}}$ is the arbitrary phase factor for the transformation which we choose to be zero such that $u_{\pmb{q}}$ are real positive. With this choice, $v_{\pmb{q}}$ are in general complex with real $v_{\pmb{0}}$. We further note that $v_{\pmb{q}} = v_{-\pmb{q}}$ and $u_{\pmb{q}} = u_{-\pmb{q}}$.

If the dipolar interaction is disregarded, $B_{\pmb{q}} = 0$ and magnons are the eigenstates of the magnetic subsystem. However, the Bogoliubov transformation necessitated by the dipolar fields leads to squeezing~\cite{Walls2008} in the magnon eigenspace giving rise to new excitations - {\em squeezed-magnons}~\cite{Kamra2016}. In the classical domain, the effect of squeezing is tantamount to an elliptical polarization of the magnons. However, the direct mathematical analogy between the squeezing of magnons and photons~\cite{Kamra2016} allows extension of the quantum effects, such as reduced vacuum fluctuations in one quadrature at the expense of the other and entanglement between different modes, already well studied for optical fields to our magnetic system. Furthermore, the expectation value of the total spin z-component:
\begin{align}
\int_{V_{\mathrm{F}}} \langle  \tilde{S}_{\mathrm{F}}^z (\pmb{r}) \rangle d^3 r & = - \frac{\mathcal{M}_0}{|\gamma|} + \sum_{\pmb{q}} \hbar (1 + 2 |v_{\pmb{q}}|^2 ) n_{\pmb{q}}^{\beta}  + \sum_{\pmb{q}} \hbar |v_{\pmb{q}}|^2 ,
\end{align}
suggests that the s-magnons possess a non-integer effective spin of $\hbar (1 + 2 |v_{\pmb{q}}|^2)$. Here, $n_{\pmb{q}}^{\beta}$ denotes the number of squeezed-magnons (s-magnons) with wavevector $\pmb{q}$, $\mathcal{M}_0 = M_s V_{\mathrm{F}}$ is the total magnetic moment, and the last term in the equation above represents vacuum noise due to squeezing~\cite{Holstein1940,Walls2008}.

\begin{figure}[htb]
\begin{center}
\subfloat[]{\includegraphics[width=75mm]{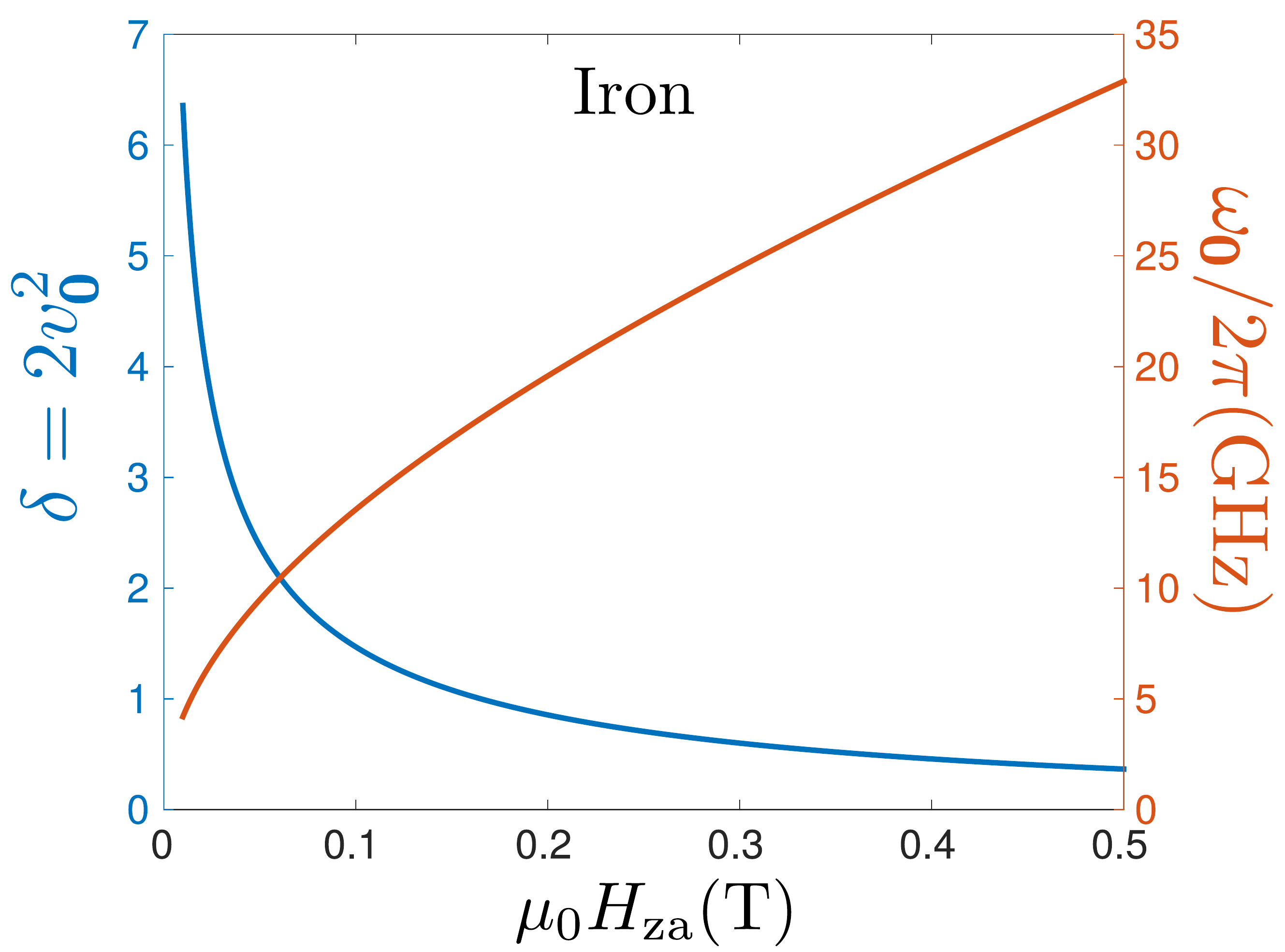}}
\subfloat[]{\includegraphics[width=75mm]{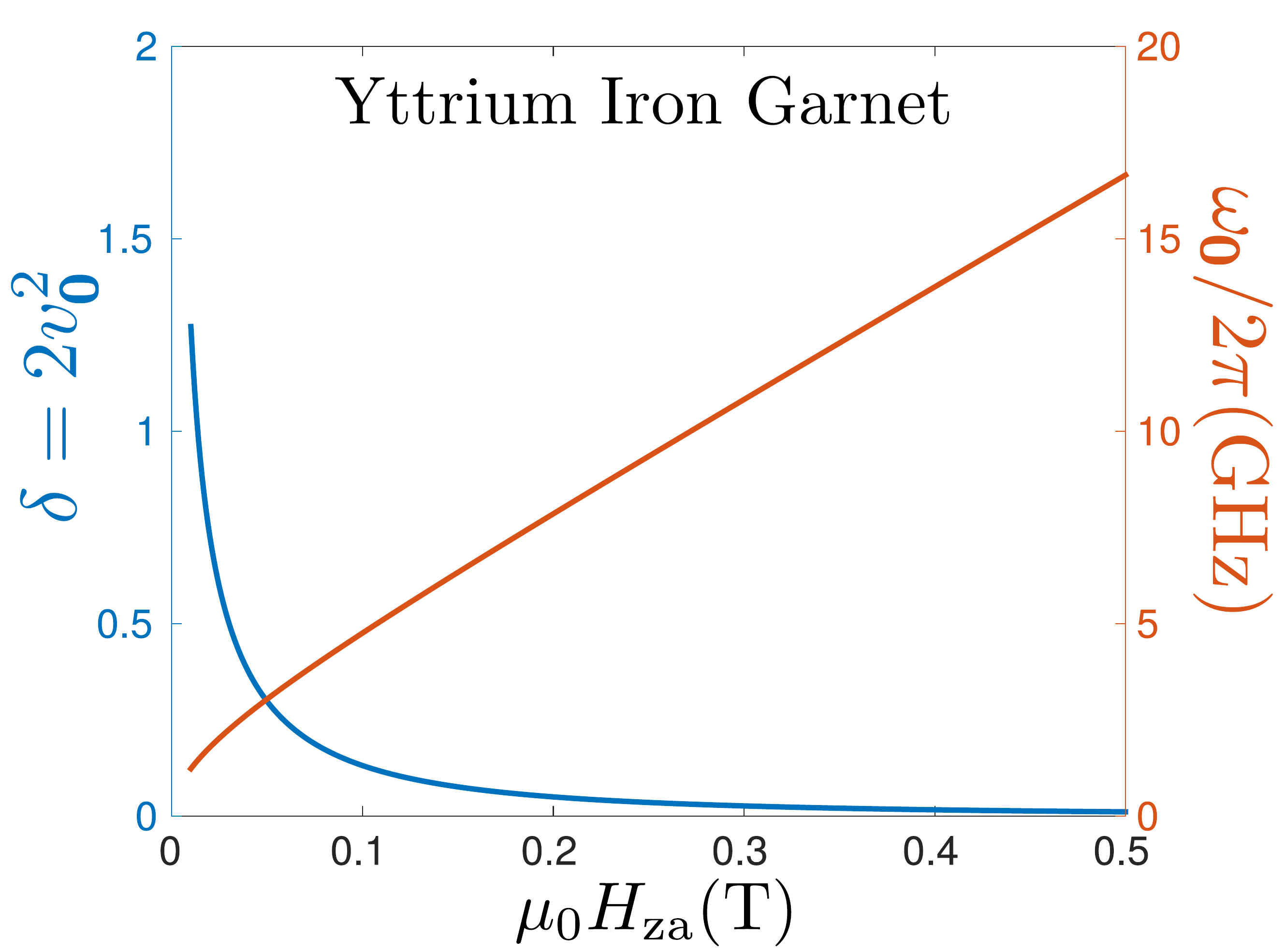}}
\caption{The eigenfrequency ($\omega_{\pmb{0}}/2 \pi$) and the squeezing mediated relative change in the effective spin ($\delta$) of the uniform s-magnon mode {\it vs.} the external plus effective anisotropy field $\mu_0 H_{za} = \omega_{za}/|\gamma|$. The degree of squeezing is larger for iron ($M_s = 1.7 \times 10^6$ A$/$m) film as compared to the YIG ($M_s = 1.4 \times 10^5$ A$/$m) film due to the former's larger saturation magnetization, and hence stronger dipolar interaction. $|\gamma| \approx 1.8 \times 10^{11}$ Hz$/$T for both materials.}
\label{fig:delnomegavsH}
\end{center}
\end{figure}

The effective spin of the uniform mode ($\pmb{q} = \pmb{0}$) is of particular interest because of the latter's central role in ferromagnetic resonance (FMR), and is given by $\hbar^* = \hbar (1 + 2 v_{\pmb{0}}^2 ) = \hbar (1 + \delta )$. We plot the relative change in the effective spin ($\delta$) along with the FMR frequency ($\omega_{\pmb{0}}/ 2 \pi$) for iron and YIG films ($N_x = 1, N_{y,z} = 0$) as a function of the external plus effective anisotropy field $\mu_0 H_{za} = \omega_{za}/|\gamma|$ in figure \ref{fig:delnomegavsH}. Within the typical experimental range of frequencies, $\delta \sim 1$ and the dipolar fields are found to play an important role. The extent of squeezing, however, is negligible whenever the contribution of dipolar interaction to the total eigenmode energy $\hbar \omega_{\pmb{q}}$ can be disregarded i.e. when $|B_{\pmb{q}}|/A_{\pmb{q}} \ll 1$. This is the case when either the Zeeman ($H_0/M_s \gg 1$) or the exchange ($D q^2 / \omega_s \gg 1$) energy dominates over the dipolar energy. Thus, in considering the phenomenon where the large $q$ excitations play the important role, the dipolar interactions and squeezing may be disregarded.

\subsubsection{Electronic and interaction contributions}\label{sssec:eleccont}
We directly write the electronic Hamiltonian $\tilde{\mathcal{H}}_{\mathrm{N}}$ diagonalized in terms of the fermionic creation ($\tilde{c}_{\pmb{k},s}^\dagger$) and annihilation ($\tilde{c}_{\pmb{k},s}$) operators corresponding to the spin-degenerate orbital wavefunctions $\psi_{\pmb{k}}(\pmb{r})$:
\begin{align}\label{hn}
\tilde{\mathcal{H}}_{\mathrm{N}} & = \sum_{\pmb{k},s} \hbar \omega_{\pmb{k}} \tilde{c}_{\pmb{k},s}^\dagger \tilde{c}_{\pmb{k},s},
\end{align}
with $s = \pm$ the index denoting electronic spin projection of $s \hbar / 2$ along the z-direction. The wavefunctions $\psi_{\pmb{k}}(\pmb{r})$, while being plane waves in the simplest case, capture essential details about the orbital dynamics in N. The spin density operator for the electronic system $\tilde{\pmb{S}}_{\mathrm{N}} (\pmb{r})$ then becomes:
\begin{align}\label{sn}
\tilde{\pmb{S}}_{\mathrm{N}}(\pmb{r}) & = \frac{\hbar}{2} \sum_{s,s^\prime}  \tilde{\Psi}_s^\dagger (\pmb{r}) \pmb{\sigma}_{s,s^\prime} \tilde{\Psi}_{s^\prime}(\pmb{r}),
\end{align}
where the components of $\pmb{\sigma}$ are the Pauli matrices, and $\tilde{\Psi}_s(\pmb{r}) = \sum_{\pmb{k}} \psi_{\pmb{k}} (\pmb{r}) \tilde{c}_{\pmb{k},s} $ is the operator that annihilates an electron with spin projection $s \hbar / 2$ at position $\pmb{r}$.

The coupling between the microwave drive and F is attributed to the Zeeman interaction between the former's oscillating magnetic field ($h_0 \cos \omega t ~ \hat{\pmb{x}}$) and the latter's net magnetic moment ($\tilde{\pmb{\mathcal{M}}} = \int_{V_{\mathrm{F}}} \tilde{\pmb{M}}(\pmb{r}) \ d^3 r$), considering the typical case of the microwave wavelength being much larger than the linear dimensions of F:
\begin{align}
\tilde{\mathcal{H}}_{\mathrm{drive}} & = - \mu_0 h_0 \tilde{\mathcal{M}}_x \cos \omega t, \\
     & = - \mu_0 h_0 B \cos \omega t \left( \tilde{\beta}_{\pmb{0}} + \tilde{\beta}_{\pmb{0}}^\dagger \right), \label{hdrive}
\end{align}
where we have employed equation (\ref{mxquant}) in obtaining the final form above, and defined $B \equiv (u_{\pmb{0}} + v_{\pmb{0}}) \sqrt{|\gamma| \hbar \mathcal{M}_0 / 2}$.

The interaction between F and N can be modeled via exchange between the interfacial spin densities in the two subsystems~\cite{Zhang2012,Bender2015}:
\begin{align}\label{hintbasic}
\tilde{\mathcal{H}}_{\mathrm{int}} & = - \frac{\mathcal{J}}{\hbar^2}  \int_{\mathcal{A}}  \tilde{\pmb{S}}_{\mathrm{F}}(\pmb{\varrho}) \cdot \tilde{\pmb{S}}_{\mathrm{N}} (\pmb{\varrho}) ~ d^2 \pmb{\varrho},
\end{align}
where $\mathcal{J}$ parametrizes the exchange strength, $\pmb{\varrho}$ denotes the in-plane 2D vector spanning the interface, and the integral is carried out over the interfacial area $\mathcal{A}$. This can be recast in terms of the creation and annihilation operators of the eigenmodes in F and N to obtain:
\begin{align}\label{hint}
\tilde{\mathcal{H}}_{\mathrm{int}} & =  \sum_{\pmb{k}_1 \pmb{k}_2 \pmb{q}} \hbar W_{\pmb{k}_1 \pmb{k}_2 \pmb{q}} ~ \tilde{c}^\dagger_{\pmb{k}_1 +} \tilde{c}_{\pmb{k}_2 -} \tilde{b}_{\pmb{q}}  \ + \ \mathrm{H.c.} , 
\end{align}
with $\tilde{b}_{\pmb{q}}  = u_{\pmb{q}} \tilde{\beta}_{\pmb{q}} + v_{\pmb{q}}^* \tilde{\beta}^\dagger_{-\pmb{q}}$, and
\begin{align}
 \hbar W_{\pmb{k}_1 \pmb{k}_2 \pmb{q}} & = \mathcal{J} \sqrt{\frac{M_s}{2 |\gamma| \hbar}} \int_{\mathcal{A}} d^2 \varrho ~ \psi_{\pmb{k}_1}^* (\pmb{\varrho}) \psi_{\pmb{k}_2} (\pmb{\varrho}) \phi_{\pmb{q}} (\pmb{\varrho}).
\end{align}
We have disregarded the terms that conserve the z-projected spin of F (and thus N) in equation (\ref{hint}). These terms do not contribute to the z-polarized spin exchange between F and N~\cite{Zhang2012}, and hence drop out in the following magnon-mediated spin current analysis. 

Since $\tilde{\mathcal{H}}_{\mathrm{int}}$ describes the interaction between F and N, it also defines the operator for the magnon-mediated (z-polarized) spin current injected into N as the interaction mediated time derivative of the total spin (z-component) in N:
\begin{align}
\tilde{I}_z \ = \tilde{\dot{\mathcal{S}}}_z  \ & = \frac{1}{i \hbar} \ \left[ \tilde{\mathcal{S}}_z , \tilde{\mathcal{H}}_{\mathrm{int}} \right], \\
       & = \sum_{\pmb{k}_1 \pmb{k}_2 \pmb{q}} - i \hbar W_{\pmb{k}_1 \pmb{k}_2 \pmb{q}} ~ \tilde{c}^\dagger_{\pmb{k}_1 +} \tilde{c}_{\pmb{k}_2 -} \tilde{b}_{\pmb{q}}  \ + \ \mathrm{H.c.} , \label{izop}
\end{align}
where $\tilde{\pmb{\mathcal{S}}} = \int_{V_{\mathrm{N}}} \tilde{\pmb{S}}_{\mathrm{N}} (\pmb{r}) \ d^3 r $ is the total spin operator in N, with $V_{\mathrm{N}}$ its volume. In steady state, the spin current injected into N dissipates due to spin relaxation yielding no net change in the N total spin. Here, we are only concerned with the spin current injection across the F$|$N interface and do not consider the spin dynamics in N.

\subsection{Equations of motion}\label{ssec:eom}
Having obtained the full Hamiltonian for the system [equations (\ref{htot}), (\ref{hf}), (\ref{hn}), (\ref{hdrive}), and (\ref{hint})], we proceed with studying the system dynamics working within the Heisenberg picture. Since all operators of interest can be expressed in terms of the eigenmode creation and annihilation operators, the time evolution of the latter gives a complete description of the system. The Heisenberg equations of motion read:
\begin{align}
\dot{\tilde{c}}_{\pmb{k} +} = & \ \frac{1}{i \hbar} \left[ \tilde{c}_{\pmb{k} +} , \tilde{\mathcal{H}}  \right] \ = \ - i \omega_{\pmb{k}} \tilde{c}_{\pmb{k} +} - i  \sum_{\pmb{k}_2 \pmb{q}} W_{\pmb{k} \pmb{k}_2 \pmb{q}} \ \tilde{c}_{\pmb{k}_2 -} \tilde{b}_{\pmb{q}}, \label{eomckplus} \\
\dot{\tilde{c}}_{\pmb{k} -}   = & - i \omega_{\pmb{k}} \tilde{c}_{\pmb{k} -} - i  \sum_{\pmb{k}_1 \pmb{q}} W_{\pmb{k}_1 \pmb{k} \pmb{q}}^* \ \tilde{c}_{\pmb{k}_1 +} \tilde{b}_{\pmb{q}}^\dagger, \label{eomckminus} \\
\dot{\tilde{\beta}}_{\pmb{q}}   = & - i \omega_{\pmb{q}} \tilde{\beta}_{\pmb{q}} - i \sum_{\pmb{k}_1 \pmb{k}_2}  \left( u_{\pmb{q}} W_{\pmb{k}_1 \pmb{k}_2 \pmb{q}}^* \ \tilde{c}_{\pmb{k}_2 - }^\dagger \tilde{c}_{\pmb{k}_1 +} + v_{\pmb{q}} W_{\pmb{k}_1 \pmb{k}_2 \pmb{q}} \ \tilde{c}_{\pmb{k}_1 + }^\dagger \tilde{c}_{\pmb{k}_2 -} \right) \nonumber \\ 
  & + i \frac{\mu_0 h_0 B}{\hbar} \cos \omega t ~ \delta_{\pmb{q},\pmb{0}}. \label{eombetaq}
\end{align}
We aim to obtain solution to these equations perturbatively up to the second order in the interfacial exchange parameter $\mathcal{J}$ [equation (\ref{hintbasic})], and hence $W_{\pmb{k}_1 \pmb{k}_2 \pmb{q}}$. To this end, we use the method employed by Gardiner and Collet~\cite{Gardiner1985} in deriving the input-output formalism~\cite{Walls2008} for quantum optical fields which entails the following mathematical prescription. Until a certain initial time $t_0$, F and N exist in thermal equilibrium without any mutual interaction or  driving field, such that the density matrix of the combined system is the outer-product of the F and N equilibrium density matrices, i.e. $\rho = \rho_{\mathrm{F}}^{\mathrm{eq}} \otimes \rho_{\mathrm{N}}^{\mathrm{eq}}$. At $t = t_0$, the F and N interaction ($\tilde{\mathcal{H}}_{\mathrm{int}}$) and the microwave drive ($\tilde{\mathcal{H}}_{\mathrm{drive}}$) are turned on. In the Heisenberg picture, the density matrix for the system stays the same while the operators evolve with time and get entangled. The steady state dynamics is obtained by taking the limit $t_0 \to - \infty$ in the end. Within this prescription, the general solution to equation (\ref{eomckplus}) for $t > t_0$ may be written as~\cite{Gardiner1985}:
\begin{align}\label{ckplusexp}
\tilde{c}_{\pmb{k} + } (t) = & e^{- i \omega_{\pmb{k}} (t - t_0)} \tilde{c}_{\pmb{k} +}(t_0) - i \sum_{\pmb{k}_2 \pmb{q}} W_{\pmb{k} \pmb{k}_2 \pmb{q}} \ \int_{t_0}^{t} e^{- i \omega_{\pmb{k}}(t - t^\prime)} \  \tilde{c}_{\pmb{k}_2 -}(t^\prime) \tilde{b}_{\pmb{q}}(t^\prime)  dt^\prime,  
\end{align}
where $ \tilde{c}_{\pmb{k} +}(t_0)$ is the initial value of the operator. In the equation above, the first term represents the unperturbed solution while the second term gives the effect of exchange interaction $\tilde{\mathcal{H}}_{\mathrm{int}}$. A similar expression follows for $\tilde{c}_{\pmb{k} - } (t)$ using equation (\ref{eomckminus}).

Since the microwave drives the $\pmb{q} = \pmb{0}$ mode coherently, represented by the last term on the right hand side of the {\em linear} dynamical equation [(\ref{eombetaq})] for $\tilde{\beta}_{\pmb{q}}$, we may express $\tilde{\beta}_{\pmb{0}} = \beta + (\tilde{\beta}_{\pmb{0}} - \beta)$ as the sum over the coherent part given by a c-number $\beta = \langle \tilde{\beta}_{\pmb{0}} \rangle$ and the incoherent part $\tilde{\beta}_{\pmb{0}} - \beta$. The dynamical equation for $\beta$ is obtained by taking the expectation value on both sides of equation (\ref{eombetaq}) for $\pmb{q} = \pmb{0}$:
\begin{align}\label{betadyn1}
\dot{\beta}  = & - i \omega_{\pmb{0}} \beta - i \sum_{\pmb{k}_1 \pmb{k}_2}  \left( u_{\pmb{0}} W_{\pmb{k}_1 \pmb{k}_2 \pmb{0}}^* \ Y_{\pmb{k}_1 \pmb{k}_2} + v_{\pmb{0}} W_{\pmb{k}_1 \pmb{k}_2 \pmb{0}} \  Y_{\pmb{k}_1 \pmb{k}_2}^* \right) + i \frac{\mu_0 h_0 B}{\hbar} \cos \omega t,
\end{align}
with $Y_{\pmb{k}_1 \pmb{k_2}} \equiv Y_{\pmb{k}_1 \pmb{k_2}}(t) = \langle \tilde{c}_{\pmb{k}_2 - }^\dagger(t) \tilde{c}_{\pmb{k}_1 +}(t) \rangle$. Employing equation (\ref{ckplusexp}) and analogous expressions for $\tilde{c}_{\pmb{k} - } (t)$ and $\tilde{\beta}_{\pmb{q}}(t)$, retaining terms up to the second order in $\mathcal{J}$, we obtain:
\begin{align}\label{yk1k2}
Y_{\pmb{k}_1 \pmb{k_2}}(t) = & i \pi W_{\pmb{k}_1 \pmb{k}_2 \pmb{0}} \ (n_{\pmb{k}_1} - n_{\pmb{k}_2}) \ \left[ u_{\pmb{0}} \beta(t) \delta(\omega_{\pmb{k}_1} - \omega_{\pmb{k}_2} - \omega) + v_{\pmb{0}} \beta^*(t) \delta(\omega_{\pmb{k}_1} - \omega_{\pmb{k}_2} + \omega) \right],
\end{align}
with $n_{\pmb{k}} = \langle \tilde{c}_{\pmb{k}}^\dagger(t_0) \tilde{c}_{\pmb{k}}(t_0) \rangle = f(\hbar \omega_{\pmb{k}} - \mu)$, where $f(\epsilon) = 1/[\exp(\epsilon/k_B T) + 1]$ is the Fermi function, $\mu$ is the chemical potential in N, $k_B$ is the Boltzmann constant, and $T$ is the system temperature. Employing equation (\ref{yk1k2}), equation (\ref{betadyn1}) simplifies to:
\begin{align}\label{betadyn2}
\dot{\beta}  = & - i \omega_{\pmb{0}} \beta - (u_{\pmb{0}}^2 + v_{\pmb{0}}^2) \Gamma_{\mathrm{N}} \beta + 2 u_{\pmb{0}} v_{\pmb{0}} \Gamma_{\mathrm{N}} \beta^* + i \frac{\mu_0 h_0 B}{\hbar} \cos \omega t,
\end{align}
where $\Gamma_{\mathrm{N}}$ is defined by:
\begin{align}\label{gamman}
\Gamma_{\mathrm{N}} \equiv \Gamma_{\mathrm{N}}(\omega) = & \sum_{\pmb{k}_1,\pmb{k}_2} \pi |W_{\pmb{k}_1 \pmb{k}_2 \pmb{0}}|^2 (n_{\pmb{k}_2} - n_{\pmb{k}_1}) \delta(\omega_{\pmb{k}_1} - \omega_{\pmb{k}_2} - \omega).
\end{align}
In writing equation (\ref{betadyn2}), we have employed the relation $\Gamma_{\mathrm{N}}(-\omega) = - \Gamma_{\mathrm{N}}(\omega)$. We now make two simplifying assumptions: (i) $|W_{\pmb{k}_1 \pmb{k}_2 \pmb{0}}|^2 \equiv |W_{\mu,\pmb{0}}|^2$, i.e. $W_{\pmb{k}_1 \pmb{k}_2 \pmb{0}}$ only depends on the magnitudes of $\pmb{k}_{1,2}$, and thus on the chemical potential in N, and (ii) the electronic density of states per unit volume in N - $g(\epsilon)$ - does not vary considerably over energy scales $k_B T$ and $\hbar \omega$ around $\epsilon = \mu$. With these assumptions, equation (\ref{gamman}) leads to the simplified expression $\Gamma_{\mathrm{N}} = \alpha^\prime \omega$, with $\alpha^\prime = \pi |W_{\mu,\pmb{0}}|^2 V_{\mathrm{N}}^2 \hbar^2 g^2(\mu)$.

Considering the ansatz $\beta = \beta_{+} \exp(i \omega t) + \beta_{-} \exp(- i \omega t)$ in equation (\ref{betadyn2}), we find that $|\beta_+| \ll |\beta_-|$ as long as $\alpha^\prime \ll 1$. Thus we may disregard the $\beta_{+}$ term thereby making the rotating wave approximation. Within this approximation, the dynamical equation for $\beta$ further simplifies to:
\begin{align}\label{betadyn3}
\dot{\beta}  = & - i \omega_{\pmb{0}} \beta - (u_{\pmb{0}}^2 + v_{\pmb{0}}^2) \Gamma_{\mathrm{N}} \beta + i \frac{\mu_0 h_0 B}{\hbar} \cos \omega t,
\end{align}
with solution:
\begin{align}\label{betat}
\beta(t) = \beta_{-} \ e^{- i \omega t} = & \frac{\mu_0 h_0 B}{2 \hbar} \ \frac{1}{(\omega_{\pmb{0}} - \omega) - i \Gamma_{\mathrm{N}} (u_{\pmb{0}}^2 + v_{\pmb{0}}^2)} \ e^{- i \omega t}.
\end{align}
Thus uniform s-magnon mode is resonantly excited for $\omega = \omega_{\pmb{0}}$ representing FMR.

It may be inferred from equations (\ref{betadyn3}) and (\ref{betat}) that $\Gamma_{\mathrm{N}}$ quantifies the dissipation of the uniform magnetic mode. Physically, $\Gamma_{\mathrm{N}}$ represents the rate at which the magnetic excitation decays due to its absorption by an N electron raising the latter from energy $\hbar \omega_{\pmb{k}_2}$ to $\hbar \omega_{\pmb{k}_1}$ [equation (\ref{gamman})]. Dissipation due to the baths internal to F (such as phonons, $\pmb{q} \neq \pmb{0}$ s-magnons, F electrons, impurities etc.) may be included similarly by considering the appropriate higher order terms in $\tilde{\mathcal{H}}_{\mathrm{F}}$~\cite{Gardiner1985}. The resulting dynamical equation for $\beta$ is then obtained by simply replacing $\Gamma_{\mathrm{N}}$ by $\Gamma = \Gamma_{\mathrm{N}} + \Gamma_{\mathrm{F}}$, where the exact form of $\Gamma_{\mathrm{F}}$ depends upon the details of the bath and the non-linear interaction considered. For the ongoing analysis, we consider $\Gamma_{\mathrm{F}} = \alpha_0 \omega$ analogous to our result for $\Gamma_{\mathrm{N}}$, and in consistence with the Landau-Lifshitz-Gilbert (LLG) phenomenology~\cite{Gilbert2004}.

\subsection{Noise evaluation}\label{ssec:neval}

The fluctuations in spin current may be quantified by the expectation value of their symmetrized correlation function: $\Phi(t_1,t_2) =  1/2 \left\langle  \tilde{\delta I}_z (t_1)  \tilde{\delta I}_z (t_2) +  \tilde{\delta I}_z (t_2)  \tilde{\delta I}_z (t_1) \right\rangle$,
where $\tilde{\delta I}_z = \tilde{I}_z - \langle \tilde{I}_z \rangle$ is the deviation of the spin current from its expectation value. Considering terms up to the second order in $\mathcal{J}$, we have
\begin{align}
\Phi(t_1,t_2) = & \frac{1}{2} \left\langle  \tilde{I}_z (t_1)  \tilde{I}_z (t_2) +  \tilde{I}_z (t_2)  \tilde{I}_z (t_1) \right\rangle, \nonumber \\
    = & \Re \left\langle \tilde{I}_z (t_1)  \tilde{I}_z (t_2) \right\rangle, \label{corr}
\end{align}
where the hermiticity of the spin current operator $\tilde{I}_z$ was employed in making the last simplification. The {\em single-sided}~\footnote{The ``single-sided'' (also known as ``one-sided'' or ``unilateral'') power spectral density $S(\Omega)$ is defined as twice the usual power for positive frequencies ($\Omega > 0$) and zero for negative frequencies ($\Omega < 0$). The definition is a matter of convenience so that in evaluating the total power in a signal via the Parseval theorem~\cite{Howard2004}, one needs to integrate over positive frequencies only.} noise power spectral density $S(\Omega)$ is obtained from the correlation function via the Wiener-Khintchine theorem~\cite{Howard2004} for non-stationary processes:
\begin{align}\label{somega}
S(\Omega) = & 2 \int_{-\infty}^{\infty} \left[ \lim_{\tau_0 \to \infty} \frac{1}{2 \tau_0} \int_{-\tau_0}^{\tau_0} \Phi(\tau,\tau - t) \ d \tau  \right] \ e^{i \Omega t} dt,
\end{align}
where the term in the square brackets is the auto-correlation function of the spin current, considering that the latter represents a non-stationary process~\cite{Howard2004} owing to the coherent drive. Since the spin current operator is proportional to $\mathcal{J}$, in evaluating the noise power up to the second order in $\mathcal{J}$, it suffices to employ the expressions for the eigenmode operators, such as equation (\ref{ckplusexp}), disregarding $\mathcal{J}$ altogether. 

Employing equations of motion for the eigenmode operators in equations (\ref{corr}) and (\ref{somega}), the noise  power conveniently separates into non-equilibrium and equilibrium contributions $S(\Omega) = S_{\mathrm{neq}}(\Omega) + S_{\mathrm{eq}}(\Omega)$. The former contribution is given by:
\begin{align}
S_{\mathrm{neq}}(\Omega) = & 2 (u_{\pmb{0}}^2 + v_{\pmb{0}}^2) \pi \hbar^2 |\beta_{-}|^2 \nonumber \\
     & \left[ h_{\pmb{0}}(\omega + \Omega) + h_{\pmb{0}}(-\omega - \Omega) + h_{\pmb{0}}(-\omega + \Omega) + h_{\pmb{0}}(\omega - \Omega) \right] , \label{sneq1}
\end{align}
where
\begin{align}\label{hx}
h_{\pmb{q}}(x) \equiv & \sum_{\pmb{k}_1,\pmb{k}_2}  |W_{\pmb{k}_1 \pmb{k}_2 \pmb{q}}|^2 n_{\pmb{k}_1} (1 - n_{\pmb{k}_2}) \delta(\omega_{\pmb{k}_1} - \omega_{\pmb{k}_2} + x).
\end{align}
The different $h_{\pmb{0}}(x)$ terms in equation (\ref{sneq1}) represent the various absorption and emission processes taking place in the system~\cite{Kamra2016}. The equilibrium noise can further be written as sum of ``classical'' [$S_{\mathrm{cl}}(\Omega)$] and ``quantum'' [$S_{\mathrm{qu}}(\Omega)$] contributions $S_{\mathrm{eq}}(\Omega) = S_{\mathrm{cl}}(\Omega) + S_{\mathrm{qu}}(\Omega)$ with:
\begin{align}
S_{\mathrm{cl}}(\Omega) = &  2 \pi \hbar^2 \sum_{\pmb{q}} (u_{\pmb{q}}^2 + |v_{\pmb{q}}|^2) n_{\pmb{q}} \nonumber \\
     & \left[ h_{\pmb{q}}(\omega_{\pmb{q}} + \Omega) + h_{\pmb{q}}(-\omega_{\pmb{q}} - \Omega) + h_{\pmb{q}}(-\omega_{\pmb{q}} + \Omega) + h_{\pmb{q}}(\omega_{\pmb{q}} - \Omega) \right] , \label{scl1} \\
S_{\mathrm{qu}}(\Omega) = &  2 \pi \hbar^2 \sum_{\pmb{q}} (u_{\pmb{q}}^2 + |v_{\pmb{q}}|^2) \left[ h_{\pmb{q}}(-\omega_{\pmb{q}} + \Omega) + h_{\pmb{q}}(-\omega_{\pmb{q}} - \Omega)  \right], \label{squ1}
\end{align}
where $n_{\pmb{q}} = n_{B}(\hbar \omega_{\pmb{q}}) \equiv 1/[\exp(\hbar \omega_{\pmb{q}}/k_B T) - 1]$ is the number of thermal s-magnons with wavevector $\pmb{q}$. The ``quantum'' contribution to noise [$S_{\mathrm{qu}}(\Omega)$] is so called since it is a direct consequence of the matrix element between two s-magnon number states being $n_{\pmb{q}} + 1$ instead of $n_{\pmb{q}}$, which is also the reason $S_{\mathrm{qu}}(\Omega)$ does not vanish at zero temperature. 

For the remaining part of this manuscript, we replace $|W_{\pmb{k}_1 \pmb{k}_2 \pmb{q}}|^2$ with $|W_{\mu,\pmb{q}}|^2$, and assume that the N electronic density of states is fairly constant around the chemical potential, analogous to the assumptions made to obtain a simple expression for $\Gamma_{\mathrm{N}}$ [equation (\ref{gamman})]. With these simplifying assumptions, equation (\ref{hx}) leads to:
\begin{align}
h_{\pmb{q}}(x) = & \hbar V_{\mathrm{N}}^2 g^2(\mu)  |W_{\mu,\pmb{q}}|^2 \ \frac{\hbar x}{1 - e^{- \frac{\hbar x}{k_B T}}}, 
\end{align}
whence, the spin current noise expressions [equations (\ref{sneq1}), (\ref{scl1}), and (\ref{squ1})] simplify to:
\begin{align}
S_{\mathrm{neq}}(\Omega) = & 2 (u_{\pmb{0}}^2 + v_{\pmb{0}}^2) \hbar \alpha^\prime |\beta_{-}|^2  \ \left[ w(\omega + \Omega) + w(\omega - \Omega)\right], \label{sneq2} \\
S_{\mathrm{cl}}(\Omega) = & \sum_{\pmb{q}} 2 \hbar (u_{\pmb{q}}^2 + |v_{\pmb{q}}|^2) \alpha^\prime_{\pmb{q}} n_{\pmb{q}} \ \left[ w(\omega_{\pmb{q}} + \Omega) + w(\omega_{\pmb{q}} - \Omega)\right], \label{scl2} \\
S_{\mathrm{qu}}(\Omega) = & \sum_{\pmb{q}} 2 \hbar^2 (u_{\pmb{q}}^2 + |v_{\pmb{q}}|^2) \alpha^\prime_{\pmb{q}} \nonumber \\
     &  \left[ (\omega_{\pmb{q}} + \Omega) n_{B}(\hbar \{\omega_{\pmb{q}} + \Omega \}) + (\omega_{\pmb{q}} - \Omega) n_{B}(\hbar \{ \omega_{\pmb{q}} - \Omega \} )  \right], \label{squ2}
\end{align}
with $w(x) \equiv \hbar x \coth (\hbar x / 2 k_B T)$, and $\alpha^\prime_{\pmb{q}} \equiv \pi |W_{\mu,\pmb{q}}|^2 V_{\mathrm{N}}^2 \hbar^2 g^2(\mu)$. Equations (\ref{sneq2}) to (\ref{squ2}) constitute the main result of this subsection.

\section{Results}\label{sec:results}

The spin current across the F$|$N  interface and its noise separates into driven (non-equilibrium) and thermal (equilibrium) contributions, with the former also depending on the temperature. We define the {\em normalized} spin current noise power, denoted by lowercase letters,  $s(\Omega) = S(\Omega)/\mathcal{A} \hbar^2 \omega_s $ as a dimensionless quantity per unit area. $s(\Omega)$ {\em approximately} represents the number of s-magnons which, if traverse unit area of the F$|$N interface every $1/\omega_s$ seconds on an average, will lead to the spin current noise $S(\Omega)$.

\subsection{Non-equilibrium}\label{ssec:neqres}

The expectation value of the net spin current is obtained from equations (\ref{izop}), (\ref{yk1k2}), and (\ref{gamman}):
\begin{align}
I_z (t) = \left\langle \tilde{I}_{z} (t) \right\rangle = I_{\mathrm{dc}} = & 2 \hbar \alpha^\prime \omega |\beta_{-}|^2,
\end{align}
employing which the spin current shot noise [equation (\ref{sneq2})] may be rewritten as:
\begin{align}\label{sneq3}
S_{\mathrm{neq}}(\Omega) = & \frac{\hbar^* I_{\mathrm{dc}}}{\hbar \omega}  \left[ w(\omega + \Omega) + w(\omega - \Omega)\right].
\end{align}
Thus $I_{\mathrm{dc}}$, and hence the shot noise, is largest under FMR $\omega = \omega_{\pmb{0}}$. In the limit of $k_B T \ll  (\hbar \omega, \hbar \Omega$), $w(x) \to \hbar |x|$ thereby recovering the result for spin current shot noise at zero temperature~\cite{Kamra2016}. The resulting zero frequency shot noise in the low temperature limit ($2 \hbar^* I_{\mathrm{dc}}$) is representative of a Poissonian spin transfer process in lumps of $\hbar^*$~\cite{Blanter2000,Kamra2016}. Thus the spin current shot noise reaffirms the non-integer spin $\hbar^*$ of the $\pmb{q} = \pmb{0}$ s-magnon mode. 

\begin{figure}[tb]
\begin{center}
\includegraphics[width=85mm]{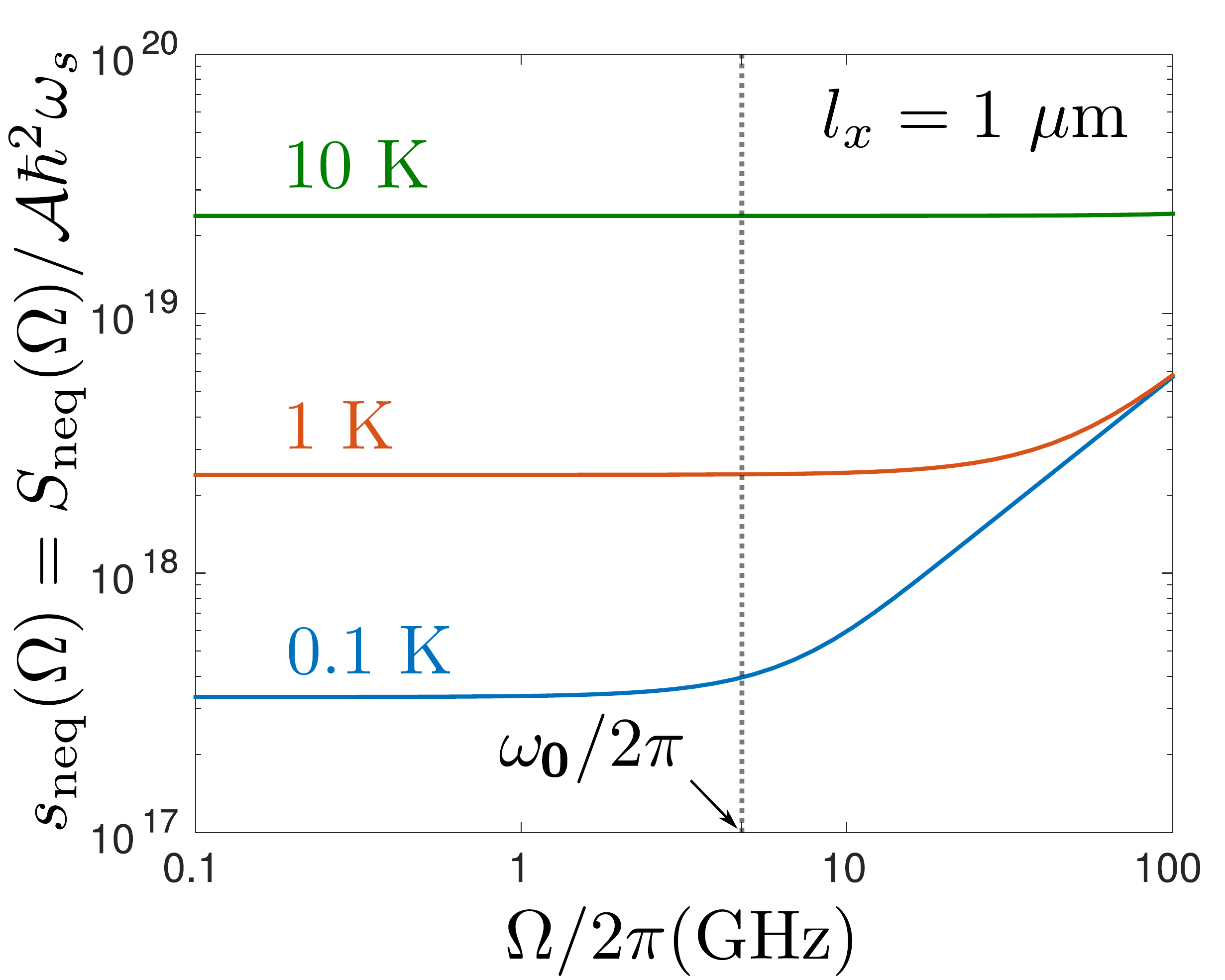}
\caption{Normalized spin current shot noise power spectra [equation (\ref{sneq3})]. The system considered is a YIG$|$Pt bilayer driven with a coherent microwave drive at ferromagnetic resonance, i.e. $\omega = \omega_{\pmb{0}}$. $l_x$ denotes the thickness of the YIG layer. }
\label{fig:Sneqvsfreq}
\end{center}
\end{figure}

On the other hand, in the high temperature limit, we obtain:
\begin{align}
S_{\mathrm{neq}}(\Omega) = & \ 2 \hbar^* I_{\mathrm{dc}} \frac{2 k_B T }{\hbar \omega}  , \quad k_B T \gg  (\hbar \omega, \hbar \Omega).
\end{align} 
Thus, in contrast with the typical situation for electronic transport~\cite{Blanter2000}, finite temperature is advantageous for measuring the magnon-mediated spin current shot noise. This difference comes about because, for the case at hand, the magnitude of $I_{\mathrm{dc}}$ is primarily determined by the microwave field amplitude $h_0$ (assuming operation under FMR), and the $2 k_B T / \hbar \omega$ enhancement is enabled by the relatively low drive frequency around FMR, $\omega \approx \omega_{\pmb{0}}$. An analogous thermal enhancement for electronic transport will require applying very low drive voltage, which in turn diminishes $I_{\mathrm{dc}}$. The (normalized) shot noise spectra [equation (\ref{sneq3})] at three different temperatures are plotted in figure \ref{fig:Sneqvsfreq} for a YIG$|$Pt bilayer with YIG thickness of 1 $\mu$m. The parameters employed in the plot are: $\omega_{\mathrm{za}} = |\gamma| \times 0.1$ T, $M_s = 1.4 \times 10^{5}$ A$/$m, $\alpha_0 = 0.001$, $|\gamma| = 1.8 \times 10^{11}$ Hz$/$T, and $\mu_0 h_0 = 100~\mu$T. Furthermore, for YIG$|$Pt bilayers, $\alpha^\prime \approx 0.215 / [l_x$ (nm)]~\cite{Tserkovnyak2002,Czeschka2011}, where $l_x$ denotes the thickness of the YIG layer. The power spectra are found to be white up to the larger between the drive frequency and $k_B T / \hbar$.

\subsection{Equilibrium}\label{ssec:eqres}

The expressions for the thermal spin current noise [equations (\ref{scl2}) and (\ref{squ2})] involve sum over all s-magnon $\pmb{q}$ modes. However, there always is an effective upper frequency cut-off, denoted here by $\omega_c$, due to the temperature or $\Omega$, which limits the number of non-vanishing terms in the sum. Furthermore, experimental data on the magnetic field dependence of the spin Seebeck effect~\cite{Boona2014,Kikkawa2015,Ritzmann2015} in the system under consideration suggests a cut-off around $\hbar \omega_{c} \approx k_B (30$ K). This latter cut-off is in addition to the analysis pursued herein. For simplicity, we make the assumption, which will be examined in detail elsewhere, $\alpha^\prime_{\pmb{q}} = \alpha^\prime_{\pmb{0}} = \alpha^\prime$. This assumption is bound to fail at large enough $\pmb{q}$ but it is acceptable for frequencies below our largest cut-off. 

In the given form, it is not possible to simplify equations (\ref{scl2}) and (\ref{squ2}) any further. We thus evaluate the noise contributions numerically and label the result with a superscript ``n''. For example, the numerically evaluated data for equation (\ref{scl2}) is denoted by $S^{\mathrm{n}}_{\mathrm{cl}}(\Omega)$, and so on. However, if we disregard dipolar interactions, simple analytical expressions for the noise power can be obtained in certain limits. We first define and discuss the validity of these limiting cases. As was discussed in section \ref{ssec:hamiltonian}, dipolar interactions play an important role for s-magnons with frequencies less than or comparable to $\omega_s$. However, the interaction may be disregarded when the dominant contribution to the thermal spin current noise comes from larger frequencies. Thus the ensuing analysis is valid when $\omega_c \gg \omega_s$. 

\begin{figure}[tb]
\begin{center}
\subfloat[]{\includegraphics[width=75mm]{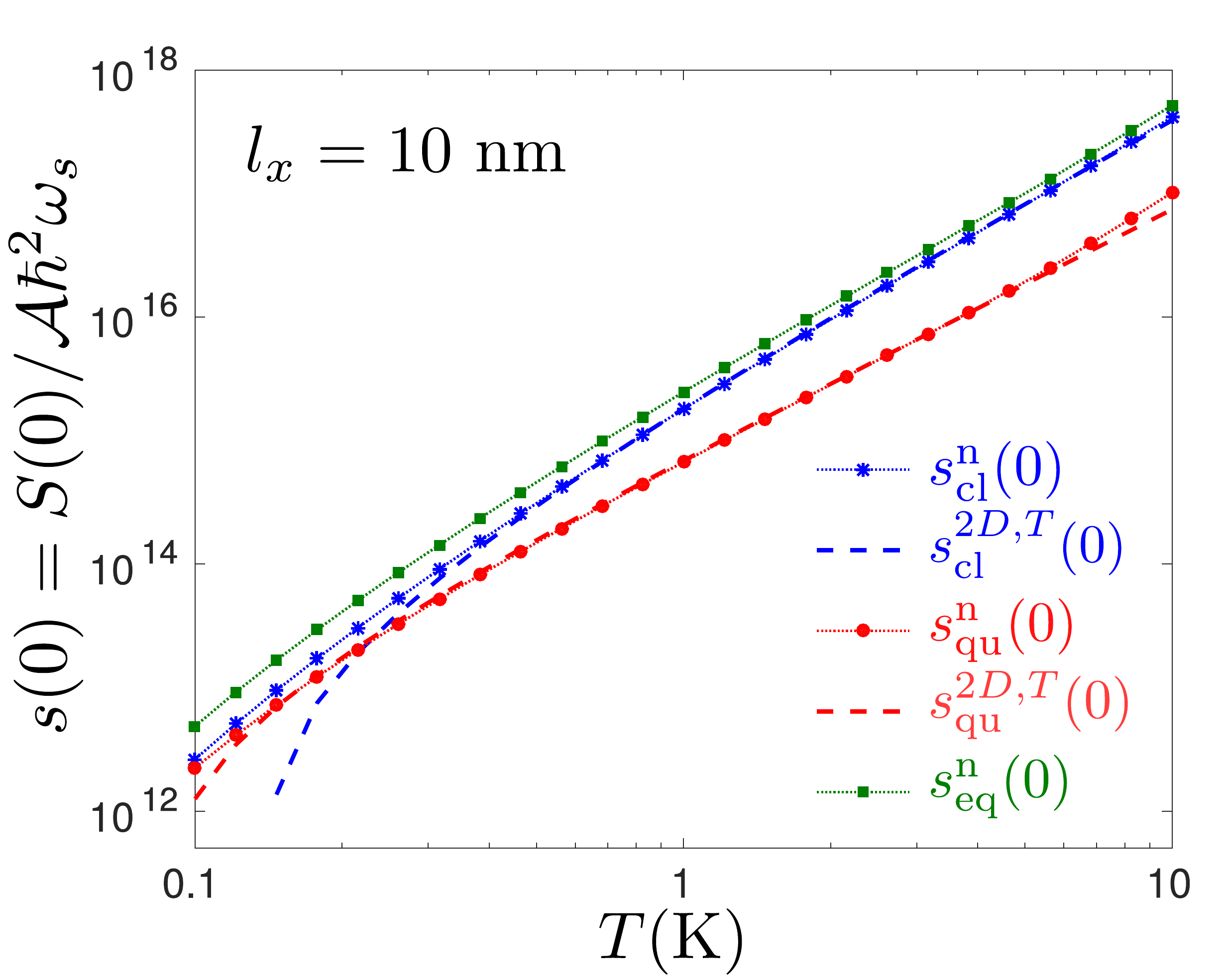}}
\subfloat[]{\includegraphics[width=75mm]{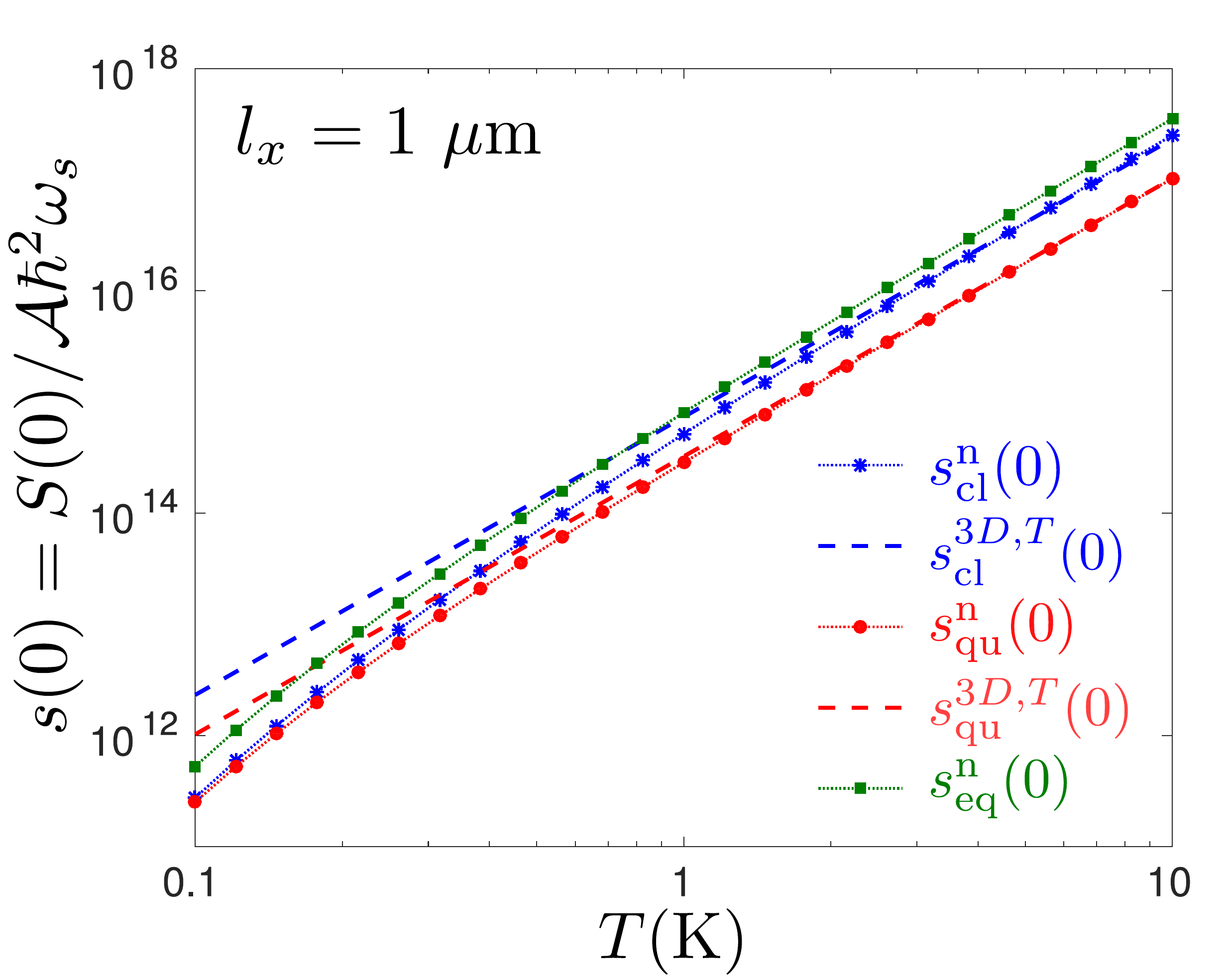}}
\caption{Normalized zero frequency noise power {\it vs.} temperature for YIG$|$Pt bilayers. The numerically evaluated results, depicted by marked-dotted lines, are compared with the analytical expressions, depicted by dashed lines. The YIG thicknesses considered are (a) 10 nm and (b) 1 $\mu$m. The former corresponds to a quasi-2D continuum while the latter to quasi-3D.}
\label{fig:SzerovsT}
\end{center}
\end{figure}

The first step towards evaluating the sum over $\pmb{q}$ is transforming it to an integral over a quasi-continuous wavevector space. The s-magnon system, however, is quasi-2D if $D/l_x^2 \gg \omega_c$ and it is quasi-3D for $D/l_x^2 \ll \omega_c$. In the following, we indicate the effective dimensionality of the magnetic subsystem by an appropriate superscript in the noise expressions. Furthermore, different expressions for noise power are obtained in the limiting cases of temperature being much larger or smaller than $\Omega$ (in the appropriate units). The larger between the two decides our effective cut-off $\omega_c$, and is thus also indicated in the superscript of the noise expressions. With these notational conventions and validity regimes, we directly write the noise expressions obtained after simplifying equations (\ref{scl2}) and (\ref{squ2}) in the quasi-2D limit:
\begin{align}
S_{\mathrm{cl}}^{\mathrm{2D,T}} (\Omega) \approx & \frac{2 \mathcal{A} \alpha^\prime k_B^2 T^2}{\pi D} \ \log \left( \frac{k_B T}{\hbar \omega_{\mathrm{za}}} \right)  \label{scl2dt},\\
S_{\mathrm{qu}}^{\mathrm{2D,T}} (\Omega) = & \frac{\mathcal{A} \alpha^\prime k_B^2 T^2}{\pi D} \ \frac{\pi^2}{6}, \label{squ2dt} \\
S_{\mathrm{cl}}^{\mathrm{2D,\Omega}} (\Omega) \approx & \frac{\mathcal{A} \hbar \alpha^\prime  k_B T}{ \pi D} \ \log \left( \frac{k_B T}{\hbar \omega_{\mathrm{za}}} \right) \ |\Omega|, \label{scl2do} \\
S_{\mathrm{qu}}^{\mathrm{2D,\Omega}} (\Omega) = & \frac{\mathcal{A} \alpha^\prime}{4 \pi D} \ (\hbar \Omega - \hbar \omega_{\mathrm{za}})^2 \ \Theta(\Omega - \omega_{\mathrm{za}}), \label{sq2do}
\end{align}
where, $\Theta(x)$ is the heaviside step function. In the quasi-3D limit:
\begin{align}
S_{\mathrm{cl}}^{\mathrm{3D,T}} (\Omega) \approx & \frac{4 V_{\mathrm{F}} \hbar \alpha^\prime }{\pi^2 (\hbar D)^{\frac{3}{2}}} \ (k_B T)^{\frac{5}{2}} , \label{scl3dt} \\
S_{\mathrm{qu}}^{\mathrm{3D,T}} (\Omega) = & \Gamma \left( 5/2 \right) \zeta \left( 5/2 \right) \frac{ V_{\mathrm{F}} \hbar \alpha^\prime }{\pi^2 (\hbar D)^{\frac{3}{2}}} \ (k_B T)^{\frac{5}{2}} , \label{squ3dt} \\
S_{\mathrm{cl}}^{\mathrm{3D,\Omega}} (\Omega) \approx & \Gamma \left( 3/2 \right) \zeta \left( 3/2 \right) \frac{ V_{\mathrm{F}} \hbar^2 \alpha^\prime }{\pi^2 (\hbar D)^{\frac{3}{2}}} \ (k_B T)^{\frac{3}{2}} \ |\Omega| , \label{scl3do} \\
S_{\mathrm{qu}}^{\mathrm{3D,\Omega}} (\Omega) = & \frac{2 V_{\mathrm{F}} \hbar \alpha^\prime }{15 \pi^2 (\hbar D)^{\frac{3}{2}}} \ (\hbar \Omega - \hbar \omega_{\mathrm{za}})^{\frac{5}{2}} \ \Theta(\Omega - \omega_{\mathrm{za}}), \label{squ3do}
\end{align}
where $\Gamma(x)$ and $\zeta(x)$ are, respectively, Gamma and Riemann Zeta functions, and the $\approx$ sign in the expressions for $S_{\mathrm{cl}} (\Omega)$ signifies that further approximations, as discussed in the appendix, have been made to obtain these closed form expressions. 

\begin{figure}[tb]
\begin{center}
\subfloat[]{\includegraphics[width=75mm]{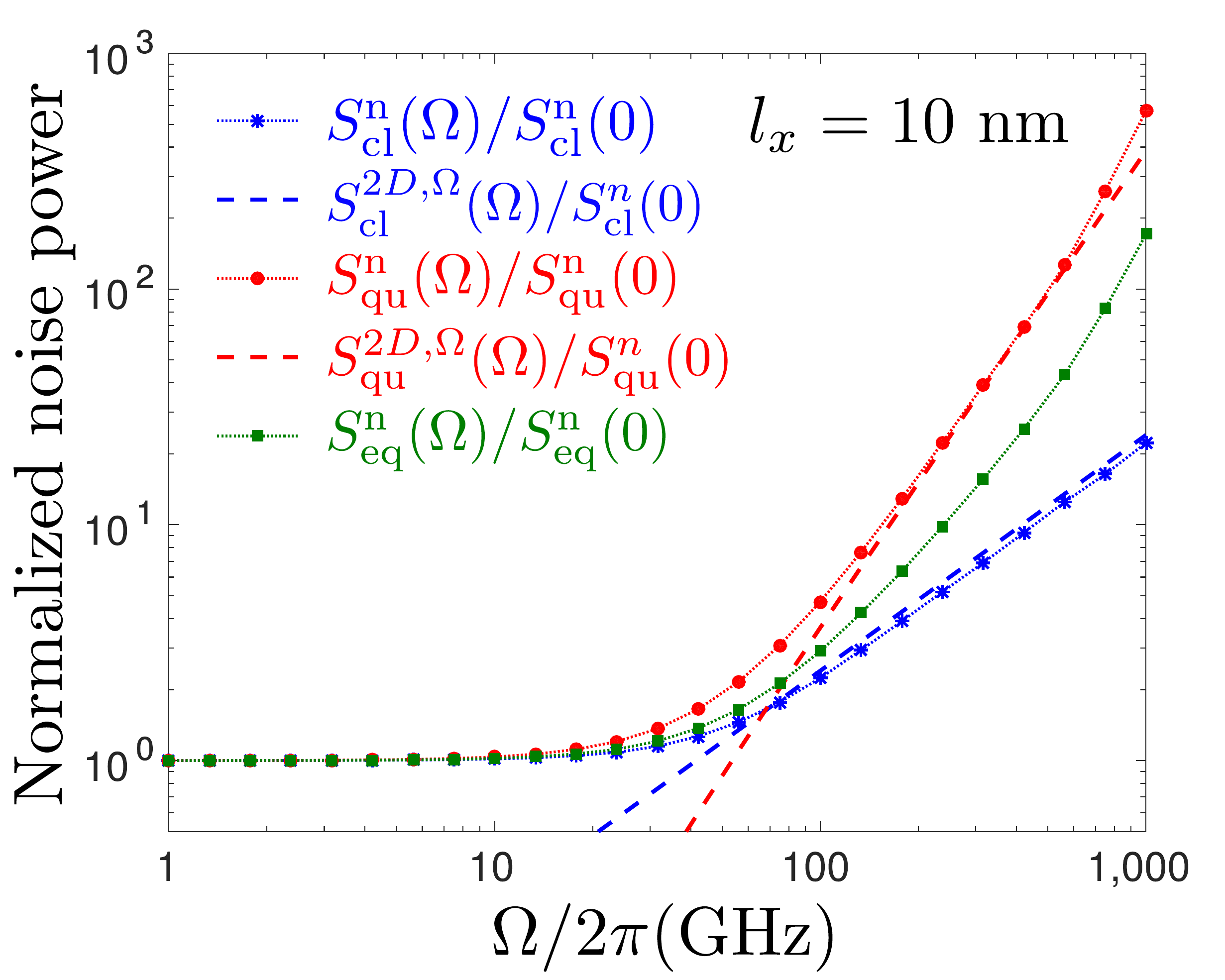}}
\subfloat[]{\includegraphics[width=75mm]{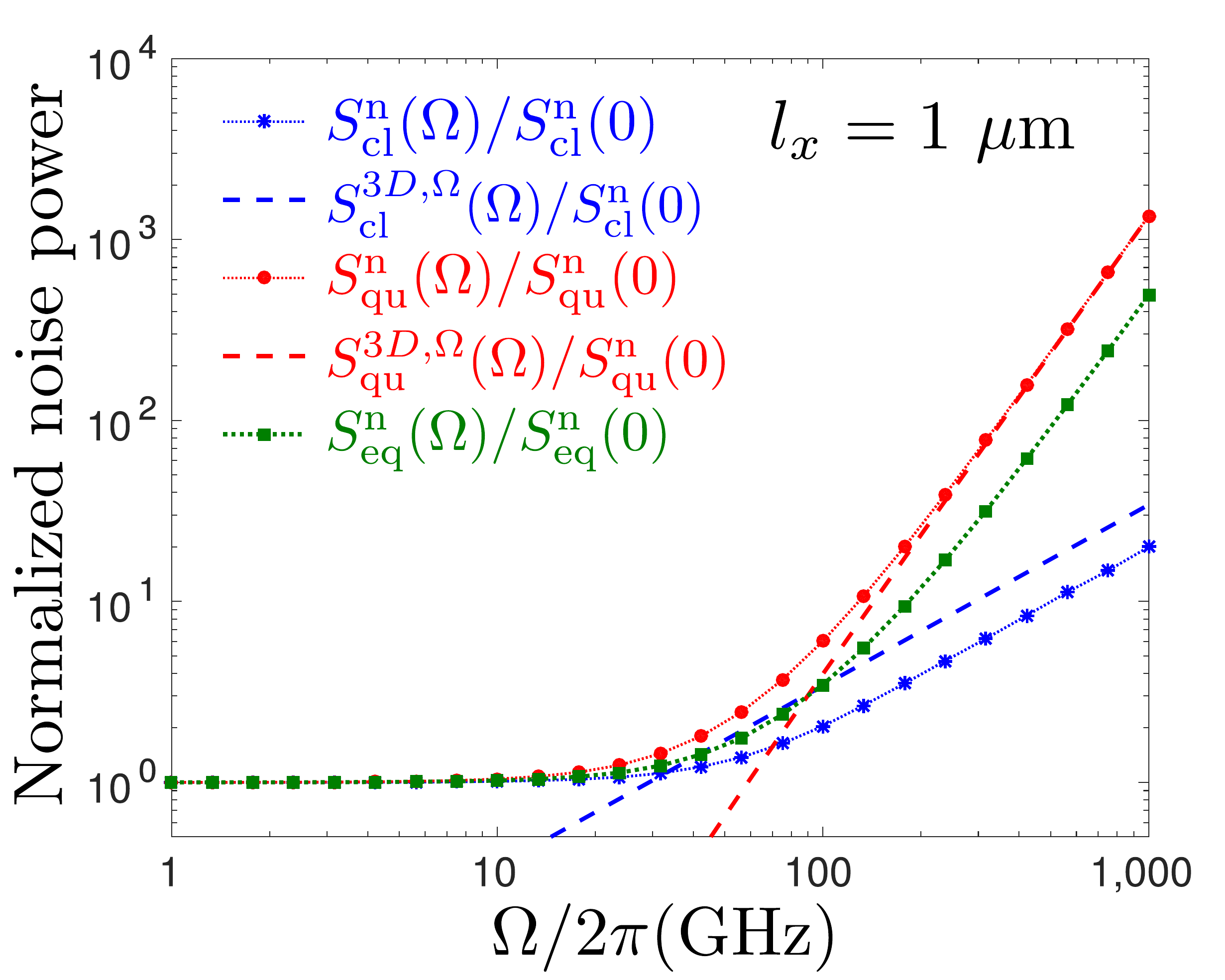}}
\caption{Noise power spectra normalized to their respective zero frequency values for YIG$|$Pt bilayers at T = 1 K. The numerically evaluated results, depicted by marked-dotted lines, are compared with the analytical expressions, depicted by dashed lines. The YIG thicknesses considered are (a) 10 nm and (b) 1 $\mu$m. The former corresponds to a quasi-2D continuum while the latter to quasi-3D.}
\label{fig:Svsfreq}
\end{center}
\end{figure}

In figure (\ref{fig:SzerovsT}), we plot the normalized zero frequency noise power {\it vs.} temperature for two different thicknesses of the YIG layer in its heterostructure with Pt. The classical and quantum contributions to the noise are comparable at very low temperatures with the former dominating as the temperature increases. In figure (\ref{fig:Svsfreq}), the frequency dependence of the noise power at a temperature of 1 K is plotted for the same bilayers. The noise power is white up to about $k_B T / \hbar$ and the quantum contribution to the noise dominates at high frequencies. The slight offsets between the numerical evaluation and analytical expressions for the classical noise stems from the crude approximations, discussed in the appendix, made in obtaining the closed form expressions. In both figures (\ref{fig:SzerovsT}) and (\ref{fig:Svsfreq}), depending on the YIG thickness, the quasi-2D (for 10 nm) or quasi-3D (for 1 $\mu$m) analytical expressions for the noise power are found to be in good agreement with the numerically evaluated results within the validity regime of the former. The parameters employed in plotting figures (\ref{fig:SzerovsT}) and (\ref{fig:Svsfreq}) are the same as those used in figure (\ref{fig:Sneqvsfreq}) with the addition: $D = 8.2 \times 10^{-6}$ $\mathrm{m}^{2} /$s~\cite{Kamra2015}.

\section{Discussion}\label{sec:discussion}

Comparing figures (\ref{fig:Sneqvsfreq}) and (\ref{fig:SzerovsT}), we find that the spin current shot noise far exceeds the thermal noise over the considered temperature range. Furthermore, due to the empirically postulated cut-off $\hbar \omega_{c} \approx k_B (30$ K) discussed above, it may be possible that the shot noise dominates the thermal noise all the way up to the room temperature. This feature is in sharp contrast with the frequency-temperature range in which the shot noise dominates in electronic (fermionic) systems, and may be understood as a special property of a non-conserved Boson gas with a coherently driven mode. Roughly speaking, the noise is directly proportional to the number of excitations. The coherent drive creates a large population of excitations in one mode while the remaining modes are populated gradually with increasing temperature. Thus, as far as the spin current across the interface is concerned, thermal noise does not pose any serious challenges to the detection of the shot noise.

Our analysis above has been perturbative in the exchange interaction between F and N. It has been tacitly assumed that the spin current exchange between F and N does not affect the distribution functions of the normal excitations in either of the subsystems. In other words, it has been assumed that the rate of equilibration in F and N far exceeds the rate of mutual spin exchange. While this is a good assumption for N (such as Pt) with strong spin-flip scattering, it breaks down for sufficiently thin F necessitating consideration of higher order terms in the perturbation parameter $\mathcal{J}$. Such an analysis accounting for the ``backflow'' effects~\cite{Sharma2016}, carried out within the LLG phenomenology for a thin films YIG$|$Pt bilayer, indicates that the thermal noise is further suppressed by the higher order correction. On the other hand, the shot noise stays the same since it has little to do with equilibration in F.

Employing the fluctuation-dissipation theorem~\cite{Callen1951}, it can be shown~\cite{Sharma2016,Kamra2014} that the thermal spin current noise is directly proportional to the spin conductivity of the F$|$N interface~\cite{Xiao2015}, i.e. the spin current absorbed by F when a non-equilibrium spin chemical potential exists in N. Thus our detailed results on spin current noise [equations (\ref{scl2dt}) to (\ref{squ3do})] also provide information on the interfacial spin conductivity over a broad parameter range.

\section{Summary}\label{sec:summary}

We have evaluated shot plus thermal noise of the spin current injected into a non-magnetic conductor (N) by an adjacent ferromagnet (F) subjected to a coherent microwave drive. The focus has been on the spin transfer mediated by the collective magnetization dynamics in F, and thus spin current polarized along the equilibrium magnetization is considered. We find that the shot noise indicates Poissonian spin transfer in lumps of $\hbar^* = \hbar (1 + \delta)$ representing the non-integer spin of the uniform squeezed-magnon~\cite{Kamra2016} mode that is driven by the coherent drive. Furthermore, the shot noise increases with temperature and is white up to the larger between the drive frequency and the temperature (in units of frequency). The thermal noise is constituted by contributions that may be classified as classical and quantum, with the latter surviving even at zero temperature. At very low temperatures, both contributions are approximately equal with the classical noise dominating as the temperature increases. On the other hand, the quantum noise is found to dominate at large frequencies while the thermal noise stays white up to the temperature (in units of frequency). The shot noise is found to dominate its thermal counterpart for typical experimental parameters encouraging the former's measurement. This shall allow for a first observation of non-integer spin of the squeezed-magnons and pave the way for exploration of their further non-trivial properties.

\section*{Acknowledgments}
We thank Sanchar Sharma (Delft) for valuable discussions, and acknowledge financial support from the Alexander von Humboldt Foundation and the DFG through SFB 767 and SPP 1538.

\appendix

\section{Sum over wavevectors}
We evaluate the classical contribution to the equilibrium noise [equation (\ref{scl2})] by transforming the sum over all s-magnon modes into integral over a quasi-continuous wavevector space. The quantum contribution to noise [equation (\ref{squ2})] can be evaluated in an analogous fashion and does not require the crude approximations, to be discussed below, that are needed for obtaining closed form expressions for the classical noise. 

Ignoring dipolar interaction such that the magnon dispersion is given by $ E/\hbar = \omega_{\pmb{q}} = \omega_{\mathrm{za}} + D q^2$ and making other assumptions discussed in section \ref{ssec:eqres}, equation (\ref{scl2}) in the quasi-2D limit leads to:
\begin{align}
S_{\mathrm{cl}}^{\mathrm{2D}}(\Omega) = & \mathcal{A}  \int  N_2(E) \ 2 \hbar \alpha^\prime \ n_{B}(E) \ \left[ w\left( \frac{E}{\hbar} + \Omega \right) + w \left( \frac{E}{\hbar} - \Omega \right)\right] \ d E, \label{sclq2d}
\end{align}
where $N_2(E) = (1/4 \pi \hbar D) \ \Theta(E - \hbar \omega_{\mathrm{za}})$ is the 2D magnon density of states, and we repeat for convenience that $n_{B}(E) = 1/[\exp(E/k_B T) - 1]$ and $w(x) = \hbar x \coth (\hbar x / 2 k_B T)$. Similarly, in the quasi-3D limit:
\begin{align}
S_{\mathrm{cl}}^{\mathrm{3D}}(\Omega) = & V_{\mathrm{F}} \int  N_3(E) \ 2 \hbar \alpha^\prime \ n_{B}(E) \ \left[ w \left( \frac{E}{\hbar} + \Omega \right) + w \left( \frac{E}{\hbar} - \Omega \right)\right] \ d E, \label{sclq3d}
\end{align}
with the 3D magnon density of states $N_3(E) = \Theta(E - \hbar \omega_{\mathrm{za}}) \ \sqrt{E - \hbar \omega_{\mathrm{za}}} / [4 \pi^2 (\hbar D)^{3/2}] $. Based on equations (\ref{sclq2d}) and (\ref{sclq3d}), we may discuss the crude approximations made in obtaining the closed form expressions for the classical noise.

In evaluating $S_{\mathrm{cl}}^{\mathrm{2D,T}}(\Omega)$ from equation (\ref{sclq2d}), we assume large temperature replacing $n_{B}(E)$ by $\Theta(k_B T - E) \ k_B T/E$ and $w(x)$ by $2 k_B T$, and obtain equation (\ref{scl2dt}) on further simplification. The exactly same replacements, in addition to disregarding $\hbar \omega_{\mathrm{za}}/ k_B T$, leads from equation (\ref{sclq3d}) to our final result for $S_{\mathrm{cl}}^{\mathrm{3D,T}}(\Omega)$ presented in equation (\ref{scl3dt}). For evaluating the analogous results in the large $\Omega$ regime [equations (\ref{scl2do}) and (\ref{scl3do})], we need to make exactly the same approximations for $n_{B}(E)$, but now $w(E / \hbar \pm \Omega)$ reduces to $\hbar |\Omega|$.

\bibliography{mag_spin_noise}

\begin{thebibliography}{47}%
\makeatletter
\providecommand \@ifxundefined [1]{%
 \@ifx{#1\undefined}
}%
\providecommand \@ifnum [1]{%
 \ifnum #1\expandafter \@firstoftwo
 \else \expandafter \@secondoftwo
 \fi
}%
\providecommand \@ifx [1]{%
 \ifx #1\expandafter \@firstoftwo
 \else \expandafter \@secondoftwo
 \fi
}%
\providecommand \natexlab [1]{#1}%
\providecommand \enquote  [1]{``#1''}%
\providecommand \bibnamefont  [1]{#1}%
\providecommand \bibfnamefont [1]{#1}%
\providecommand \citenamefont [1]{#1}%
\providecommand \href@noop [0]{\@secondoftwo}%
\providecommand \href [0]{\begingroup \@sanitize@url \@href}%
\providecommand \@href[1]{\@@startlink{#1}\@@href}%
\providecommand \@@href[1]{\endgroup#1\@@endlink}%
\providecommand \@sanitize@url [0]{\catcode `\\12\catcode `\$12\catcode
  `\&12\catcode `\#12\catcode `\^12\catcode `\_12\catcode `\%12\relax}%
\providecommand \@@startlink[1]{}%
\providecommand \@@endlink[0]{}%
\providecommand \url  [0]{\begingroup\@sanitize@url \@url }%
\providecommand \@url [1]{\endgroup\@href {#1}{\urlprefix }}%
\providecommand \urlprefix  [0]{URL }%
\providecommand \Eprint [0]{\href }%
\providecommand \doibase [0]{http://dx.doi.org/}%
\providecommand \selectlanguage [0]{\@gobble}%
\providecommand \bibinfo  [0]{\@secondoftwo}%
\providecommand \bibfield  [0]{\@secondoftwo}%
\providecommand \translation [1]{[#1]}%
\providecommand \BibitemOpen [0]{}%
\providecommand \bibitemStop [0]{}%
\providecommand \bibitemNoStop [0]{.\EOS\space}%
\providecommand \EOS [0]{\spacefactor3000\relax}%
\providecommand \BibitemShut  [1]{\csname bibitem#1\endcsname}%
\let\auto@bib@innerbib\@empty
\bibitem [{\citenamefont {Tserkovnyak}\ \emph {et~al.}(2002)\citenamefont
  {Tserkovnyak}, \citenamefont {Brataas},\ and\ \citenamefont
  {Bauer}}]{Tserkovnyak2002}%
  \BibitemOpen
  \bibfield  {author} {\bibinfo {author} {\bibfnamefont {Y.}~\bibnamefont
  {Tserkovnyak}}, \bibinfo {author} {\bibfnamefont {A.}~\bibnamefont
  {Brataas}}, \ and\ \bibinfo {author} {\bibfnamefont {G.~E.~W.}\ \bibnamefont
  {Bauer}},\ }\href {\doibase 10.1103/PhysRevLett.88.117601} {\bibfield
  {journal} {\bibinfo  {journal} {Phys. Rev. Lett.}\ }\textbf {\bibinfo
  {volume} {88}},\ \bibinfo {pages} {117601} (\bibinfo {year}
  {2002})}\BibitemShut {NoStop}%
\bibitem [{\citenamefont {Weiler}\ \emph {et~al.}(2013)\citenamefont {Weiler},
  \citenamefont {Althammer}, \citenamefont {Schreier}, \citenamefont {Lotze},
  \citenamefont {Pernpeintner}, \citenamefont {Meyer}, \citenamefont {Huebl},
  \citenamefont {Gross}, \citenamefont {Kamra}, \citenamefont {Xiao},
  \citenamefont {Chen}, \citenamefont {Jiao}, \citenamefont {Bauer},\ and\
  \citenamefont {Goennenwein}}]{Weiler2013}%
  \BibitemOpen
  \bibfield  {author} {\bibinfo {author} {\bibfnamefont {M.}~\bibnamefont
  {Weiler}}, \bibinfo {author} {\bibfnamefont {M.}~\bibnamefont {Althammer}},
  \bibinfo {author} {\bibfnamefont {M.}~\bibnamefont {Schreier}}, \bibinfo
  {author} {\bibfnamefont {J.}~\bibnamefont {Lotze}}, \bibinfo {author}
  {\bibfnamefont {M.}~\bibnamefont {Pernpeintner}}, \bibinfo {author}
  {\bibfnamefont {S.}~\bibnamefont {Meyer}}, \bibinfo {author} {\bibfnamefont
  {H.}~\bibnamefont {Huebl}}, \bibinfo {author} {\bibfnamefont
  {R.}~\bibnamefont {Gross}}, \bibinfo {author} {\bibfnamefont
  {A.}~\bibnamefont {Kamra}}, \bibinfo {author} {\bibfnamefont
  {J.}~\bibnamefont {Xiao}}, \bibinfo {author} {\bibfnamefont {Y.-T.}\
  \bibnamefont {Chen}}, \bibinfo {author} {\bibfnamefont {H.}~\bibnamefont
  {Jiao}}, \bibinfo {author} {\bibfnamefont {G.~E.~W.}\ \bibnamefont {Bauer}},
  \ and\ \bibinfo {author} {\bibfnamefont {S.~T.~B.}\ \bibnamefont
  {Goennenwein}},\ }\href {\doibase 10.1103/PhysRevLett.111.176601} {\bibfield
  {journal} {\bibinfo  {journal} {Phys. Rev. Lett.}\ }\textbf {\bibinfo
  {volume} {111}},\ \bibinfo {pages} {176601} (\bibinfo {year}
  {2013})}\BibitemShut {NoStop}%
\bibitem [{\citenamefont {Kruglyak}\ \emph {et~al.}(2010)\citenamefont
  {Kruglyak}, \citenamefont {Demokritov},\ and\ \citenamefont
  {Grundler}}]{Kruglyak2010}%
  \BibitemOpen
  \bibfield  {author} {\bibinfo {author} {\bibfnamefont {V.~V.}\ \bibnamefont
  {Kruglyak}}, \bibinfo {author} {\bibfnamefont {S.~O.}\ \bibnamefont
  {Demokritov}}, \ and\ \bibinfo {author} {\bibfnamefont {D.}~\bibnamefont
  {Grundler}},\ }\href {http://stacks.iop.org/0022-3727/43/i=26/a=264001}
  {\bibfield  {journal} {\bibinfo  {journal} {Journal of Physics D: Applied
  Physics}\ }\textbf {\bibinfo {volume} {43}},\ \bibinfo {pages} {264001}
  (\bibinfo {year} {2010})}\BibitemShut {NoStop}%
\bibitem [{\citenamefont {Shindou}\ \emph {et~al.}(2013)\citenamefont
  {Shindou}, \citenamefont {Matsumoto}, \citenamefont {Murakami},\ and\
  \citenamefont {Ohe}}]{Shindou2013}%
  \BibitemOpen
  \bibfield  {author} {\bibinfo {author} {\bibfnamefont {R.}~\bibnamefont
  {Shindou}}, \bibinfo {author} {\bibfnamefont {R.}~\bibnamefont {Matsumoto}},
  \bibinfo {author} {\bibfnamefont {S.}~\bibnamefont {Murakami}}, \ and\
  \bibinfo {author} {\bibfnamefont {J.-i.}\ \bibnamefont {Ohe}},\ }\href
  {\doibase 10.1103/PhysRevB.87.174427} {\bibfield  {journal} {\bibinfo
  {journal} {Phys. Rev. B}\ }\textbf {\bibinfo {volume} {87}},\ \bibinfo
  {pages} {174427} (\bibinfo {year} {2013})}\BibitemShut {NoStop}%
\bibitem [{\citenamefont {Demokritov}\ \emph {et~al.}(2006)\citenamefont
  {Demokritov}, \citenamefont {Demidov}, \citenamefont {Dzyapko}, \citenamefont
  {Melkov}, \citenamefont {Serga}, \citenamefont {Hillebrands},\ and\
  \citenamefont {Slavin}}]{Demokritov2006}%
  \BibitemOpen
  \bibfield  {author} {\bibinfo {author} {\bibfnamefont {S.~O.}\ \bibnamefont
  {Demokritov}}, \bibinfo {author} {\bibfnamefont {V.~E.}\ \bibnamefont
  {Demidov}}, \bibinfo {author} {\bibfnamefont {O.}~\bibnamefont {Dzyapko}},
  \bibinfo {author} {\bibfnamefont {G.~A.}\ \bibnamefont {Melkov}}, \bibinfo
  {author} {\bibfnamefont {A.~A.}\ \bibnamefont {Serga}}, \bibinfo {author}
  {\bibfnamefont {B.}~\bibnamefont {Hillebrands}}, \ and\ \bibinfo {author}
  {\bibfnamefont {A.~N.}\ \bibnamefont {Slavin}},\ }\href {\doibase
  10.1038/nature05117} {\bibfield  {journal} {\bibinfo  {journal} {Nature}\
  }\textbf {\bibinfo {volume} {443}},\ \bibinfo {pages} {430} (\bibinfo {year}
  {2006})}\BibitemShut {NoStop}%
\bibitem [{\citenamefont {Brataas}\ \emph {et~al.}(2012)\citenamefont
  {Brataas}, \citenamefont {Tserkovnyak}, \citenamefont {Bauer},\ and\
  \citenamefont {Kelly}}]{Brataas2012}%
  \BibitemOpen
  \bibfield  {author} {\bibinfo {author} {\bibfnamefont {A.}~\bibnamefont
  {Brataas}}, \bibinfo {author} {\bibfnamefont {Y.}~\bibnamefont
  {Tserkovnyak}}, \bibinfo {author} {\bibfnamefont {G.~E.~W.}\ \bibnamefont
  {Bauer}}, \ and\ \bibinfo {author} {\bibfnamefont {P.~J.}\ \bibnamefont
  {Kelly}},\ }in\ \href@noop {} {\emph {\bibinfo {booktitle} {Spin Current}}},\
  \bibinfo {series and number} {Series on Semiconductor Science and
  Technology},\ \bibinfo {editor} {edited by\ \bibinfo {editor} {\bibfnamefont
  {S.}~\bibnamefont {Maekawa}}, \bibinfo {editor} {\bibfnamefont
  {S.}~\bibnamefont {Valenzuela}}, \bibinfo {editor} {\bibfnamefont
  {E.}~\bibnamefont {Saitoh}}, \ and\ \bibinfo {editor} {\bibfnamefont
  {T.}~\bibnamefont {Kimura}}}\ (\bibinfo  {publisher} {Oxford University
  Press},\ \bibinfo {address} {Oxford},\ \bibinfo {year} {2012})\BibitemShut
  {NoStop}%
\bibitem [{\citenamefont {Uchida}\ \emph {et~al.}(2010)\citenamefont {Uchida},
  \citenamefont {Xiao}, \citenamefont {Adachi}, \citenamefont {Ohe},
  \citenamefont {Takahashi}, \citenamefont {Ieda}, \citenamefont {Ota},
  \citenamefont {Kajiwara}, \citenamefont {Umezawa}, \citenamefont {Kawai},
  \citenamefont {Bauer}, \citenamefont {Maekawa},\ and\ \citenamefont
  {Saitoh}}]{Uchida2010}%
  \BibitemOpen
  \bibfield  {author} {\bibinfo {author} {\bibfnamefont {K.}~\bibnamefont
  {Uchida}}, \bibinfo {author} {\bibfnamefont {J.}~\bibnamefont {Xiao}},
  \bibinfo {author} {\bibfnamefont {H.}~\bibnamefont {Adachi}}, \bibinfo
  {author} {\bibfnamefont {J.}~\bibnamefont {Ohe}}, \bibinfo {author}
  {\bibfnamefont {S.}~\bibnamefont {Takahashi}}, \bibinfo {author}
  {\bibfnamefont {J.}~\bibnamefont {Ieda}}, \bibinfo {author} {\bibfnamefont
  {T.}~\bibnamefont {Ota}}, \bibinfo {author} {\bibfnamefont {Y.}~\bibnamefont
  {Kajiwara}}, \bibinfo {author} {\bibfnamefont {H.}~\bibnamefont {Umezawa}},
  \bibinfo {author} {\bibfnamefont {H.}~\bibnamefont {Kawai}}, \bibinfo
  {author} {\bibfnamefont {G.~E.~W.}\ \bibnamefont {Bauer}}, \bibinfo {author}
  {\bibfnamefont {S.}~\bibnamefont {Maekawa}}, \ and\ \bibinfo {author}
  {\bibfnamefont {E.}~\bibnamefont {Saitoh}},\ }\href {\doibase
  10.1038/nmat2856} {\bibfield  {journal} {\bibinfo  {journal} {Nat Mater}\
  }\textbf {\bibinfo {volume} {9}},\ \bibinfo {pages} {894} (\bibinfo {year}
  {2010})}\BibitemShut {NoStop}%
\bibitem [{\citenamefont {Xiao}\ \emph {et~al.}(2010)\citenamefont {Xiao},
  \citenamefont {Bauer}, \citenamefont {Uchida}, \citenamefont {Saitoh},\ and\
  \citenamefont {Maekawa}}]{Xiao2010}%
  \BibitemOpen
  \bibfield  {author} {\bibinfo {author} {\bibfnamefont {J.}~\bibnamefont
  {Xiao}}, \bibinfo {author} {\bibfnamefont {G.~E.~W.}\ \bibnamefont {Bauer}},
  \bibinfo {author} {\bibfnamefont {K.-c.}\ \bibnamefont {Uchida}}, \bibinfo
  {author} {\bibfnamefont {E.}~\bibnamefont {Saitoh}}, \ and\ \bibinfo {author}
  {\bibfnamefont {S.}~\bibnamefont {Maekawa}},\ }\href {\doibase
  10.1103/PhysRevB.81.214418} {\bibfield  {journal} {\bibinfo  {journal} {Phys.
  Rev. B}\ }\textbf {\bibinfo {volume} {81}},\ \bibinfo {pages} {214418}
  (\bibinfo {year} {2010})}\BibitemShut {NoStop}%
\bibitem [{\citenamefont {Adachi}\ \emph {et~al.}(2013)\citenamefont {Adachi},
  \citenamefont {ichi Uchida}, \citenamefont {Saitoh},\ and\ \citenamefont
  {Maekawa}}]{Adachi2013}%
  \BibitemOpen
  \bibfield  {author} {\bibinfo {author} {\bibfnamefont {H.}~\bibnamefont
  {Adachi}}, \bibinfo {author} {\bibfnamefont {K.}~\bibnamefont {ichi Uchida}},
  \bibinfo {author} {\bibfnamefont {E.}~\bibnamefont {Saitoh}}, \ and\ \bibinfo
  {author} {\bibfnamefont {S.}~\bibnamefont {Maekawa}},\ }\href
  {http://stacks.iop.org/0034-4885/76/i=3/a=036501} {\bibfield  {journal}
  {\bibinfo  {journal} {Reports on Progress in Physics}\ }\textbf {\bibinfo
  {volume} {76}},\ \bibinfo {pages} {036501} (\bibinfo {year}
  {2013})}\BibitemShut {NoStop}%
\bibitem [{\citenamefont {Bauer}\ \emph {et~al.}(2012)\citenamefont {Bauer},
  \citenamefont {Saitoh},\ and\ \citenamefont {van Wees}}]{Bauer2012}%
  \BibitemOpen
  \bibfield  {author} {\bibinfo {author} {\bibfnamefont {G.~E.~W.}\
  \bibnamefont {Bauer}}, \bibinfo {author} {\bibfnamefont {E.}~\bibnamefont
  {Saitoh}}, \ and\ \bibinfo {author} {\bibfnamefont {B.~J.}\ \bibnamefont {van
  Wees}},\ }\href {\doibase http://dx.doi.org/10.1038/nmat3301} {\bibfield
  {journal} {\bibinfo  {journal} {Nat Mater}\ }\textbf {\bibinfo {volume}
  {11}},\ \bibinfo {pages} {391} (\bibinfo {year} {2012})}\BibitemShut
  {NoStop}%
\bibitem [{\citenamefont {Bender}\ and\ \citenamefont
  {Tserkovnyak}(2015)}]{Bender2015}%
  \BibitemOpen
  \bibfield  {author} {\bibinfo {author} {\bibfnamefont {S.~A.}\ \bibnamefont
  {Bender}}\ and\ \bibinfo {author} {\bibfnamefont {Y.}~\bibnamefont
  {Tserkovnyak}},\ }\href {\doibase 10.1103/PhysRevB.91.140402} {\bibfield
  {journal} {\bibinfo  {journal} {Phys. Rev. B}\ }\textbf {\bibinfo {volume}
  {91}},\ \bibinfo {pages} {140402} (\bibinfo {year} {2015})}\BibitemShut
  {NoStop}%
\bibitem [{\citenamefont {{Xiao}}\ and\ \citenamefont
  {{Bauer}}(2015)}]{Xiao2015}%
  \BibitemOpen
  \bibfield  {author} {\bibinfo {author} {\bibfnamefont {J.}~\bibnamefont
  {{Xiao}}}\ and\ \bibinfo {author} {\bibfnamefont {G.~E.~W.}\ \bibnamefont
  {{Bauer}}},\ }\href@noop {} {\bibfield  {journal} {\bibinfo  {journal} {ArXiv
  e-prints}\ } (\bibinfo {year} {2015})},\ \Eprint
  {http://arxiv.org/abs/1508.02486} {arXiv:1508.02486 [cond-mat.mtrl-sci]}
  \BibitemShut {NoStop}%
\bibitem [{\citenamefont {Brown}(1963)}]{Brown1963}%
  \BibitemOpen
  \bibfield  {author} {\bibinfo {author} {\bibfnamefont {W.~F.}\ \bibnamefont
  {Brown}},\ }\href {\doibase 10.1103/PhysRev.130.1677} {\bibfield  {journal}
  {\bibinfo  {journal} {Phys. Rev.}\ }\textbf {\bibinfo {volume} {130}},\
  \bibinfo {pages} {1677} (\bibinfo {year} {1963})}\BibitemShut {NoStop}%
\bibitem [{\citenamefont {Safonov}\ and\ \citenamefont
  {Bertram}(2002)}]{Safonov2002}%
  \BibitemOpen
  \bibfield  {author} {\bibinfo {author} {\bibfnamefont {V.~L.}\ \bibnamefont
  {Safonov}}\ and\ \bibinfo {author} {\bibfnamefont {H.~N.}\ \bibnamefont
  {Bertram}},\ }\href {\doibase 10.1103/PhysRevB.65.172417} {\bibfield
  {journal} {\bibinfo  {journal} {Phys. Rev. B}\ }\textbf {\bibinfo {volume}
  {65}},\ \bibinfo {pages} {172417} (\bibinfo {year} {2002})}\BibitemShut
  {NoStop}%
\bibitem [{\citenamefont {Foros}\ \emph {et~al.}(2005)\citenamefont {Foros},
  \citenamefont {Brataas}, \citenamefont {Tserkovnyak},\ and\ \citenamefont
  {Bauer}}]{Foros2005}%
  \BibitemOpen
  \bibfield  {author} {\bibinfo {author} {\bibfnamefont {J.}~\bibnamefont
  {Foros}}, \bibinfo {author} {\bibfnamefont {A.}~\bibnamefont {Brataas}},
  \bibinfo {author} {\bibfnamefont {Y.}~\bibnamefont {Tserkovnyak}}, \ and\
  \bibinfo {author} {\bibfnamefont {G.~E.~W.}\ \bibnamefont {Bauer}},\ }\href
  {\doibase 10.1103/PhysRevLett.95.016601} {\bibfield  {journal} {\bibinfo
  {journal} {Phys. Rev. Lett.}\ }\textbf {\bibinfo {volume} {95}},\ \bibinfo
  {pages} {016601} (\bibinfo {year} {2005})}\BibitemShut {NoStop}%
\bibitem [{\citenamefont {Rossi}\ \emph {et~al.}(2005)\citenamefont {Rossi},
  \citenamefont {Heinonen},\ and\ \citenamefont {MacDonald}}]{Rossi2005}%
  \BibitemOpen
  \bibfield  {author} {\bibinfo {author} {\bibfnamefont {E.}~\bibnamefont
  {Rossi}}, \bibinfo {author} {\bibfnamefont {O.~G.}\ \bibnamefont {Heinonen}},
  \ and\ \bibinfo {author} {\bibfnamefont {A.~H.}\ \bibnamefont {MacDonald}},\
  }\href {\doibase 10.1103/PhysRevB.72.174412} {\bibfield  {journal} {\bibinfo
  {journal} {Phys. Rev. B}\ }\textbf {\bibinfo {volume} {72}},\ \bibinfo
  {pages} {174412} (\bibinfo {year} {2005})}\BibitemShut {NoStop}%
\bibitem [{\citenamefont {Belzig}\ and\ \citenamefont
  {Zareyan}(2004)}]{Belzig2004}%
  \BibitemOpen
  \bibfield  {author} {\bibinfo {author} {\bibfnamefont {W.}~\bibnamefont
  {Belzig}}\ and\ \bibinfo {author} {\bibfnamefont {M.}~\bibnamefont
  {Zareyan}},\ }\href {\doibase 10.1103/PhysRevB.69.140407} {\bibfield
  {journal} {\bibinfo  {journal} {Phys. Rev. B}\ }\textbf {\bibinfo {volume}
  {69}},\ \bibinfo {pages} {140407} (\bibinfo {year} {2004})}\BibitemShut
  {NoStop}%
\bibitem [{\citenamefont {Kamra}\ \emph {et~al.}(2014)\citenamefont {Kamra},
  \citenamefont {Witek}, \citenamefont {Meyer}, \citenamefont {Huebl},
  \citenamefont {Gepr\"ags}, \citenamefont {Gross}, \citenamefont {Bauer},\
  and\ \citenamefont {Goennenwein}}]{Kamra2014}%
  \BibitemOpen
  \bibfield  {author} {\bibinfo {author} {\bibfnamefont {A.}~\bibnamefont
  {Kamra}}, \bibinfo {author} {\bibfnamefont {F.~P.}\ \bibnamefont {Witek}},
  \bibinfo {author} {\bibfnamefont {S.}~\bibnamefont {Meyer}}, \bibinfo
  {author} {\bibfnamefont {H.}~\bibnamefont {Huebl}}, \bibinfo {author}
  {\bibfnamefont {S.}~\bibnamefont {Gepr\"ags}}, \bibinfo {author}
  {\bibfnamefont {R.}~\bibnamefont {Gross}}, \bibinfo {author} {\bibfnamefont
  {G.~E.~W.}\ \bibnamefont {Bauer}}, \ and\ \bibinfo {author} {\bibfnamefont
  {S.~T.~B.}\ \bibnamefont {Goennenwein}},\ }\href {\doibase
  10.1103/PhysRevB.90.214419} {\bibfield  {journal} {\bibinfo  {journal} {Phys.
  Rev. B}\ }\textbf {\bibinfo {volume} {90}},\ \bibinfo {pages} {214419}
  (\bibinfo {year} {2014})}\BibitemShut {NoStop}%
\bibitem [{\citenamefont {Arakawa}\ \emph {et~al.}(2015)\citenamefont
  {Arakawa}, \citenamefont {Shiogai}, \citenamefont {Ciorga}, \citenamefont
  {Utz}, \citenamefont {Schuh}, \citenamefont {Kohda}, \citenamefont {Nitta},
  \citenamefont {Bougeard}, \citenamefont {Weiss}, \citenamefont {Ono},\ and\
  \citenamefont {Kobayashi}}]{Arakawa2015}%
  \BibitemOpen
  \bibfield  {author} {\bibinfo {author} {\bibfnamefont {T.}~\bibnamefont
  {Arakawa}}, \bibinfo {author} {\bibfnamefont {J.}~\bibnamefont {Shiogai}},
  \bibinfo {author} {\bibfnamefont {M.}~\bibnamefont {Ciorga}}, \bibinfo
  {author} {\bibfnamefont {M.}~\bibnamefont {Utz}}, \bibinfo {author}
  {\bibfnamefont {D.}~\bibnamefont {Schuh}}, \bibinfo {author} {\bibfnamefont
  {M.}~\bibnamefont {Kohda}}, \bibinfo {author} {\bibfnamefont
  {J.}~\bibnamefont {Nitta}}, \bibinfo {author} {\bibfnamefont
  {D.}~\bibnamefont {Bougeard}}, \bibinfo {author} {\bibfnamefont
  {D.}~\bibnamefont {Weiss}}, \bibinfo {author} {\bibfnamefont
  {T.}~\bibnamefont {Ono}}, \ and\ \bibinfo {author} {\bibfnamefont
  {K.}~\bibnamefont {Kobayashi}},\ }\href {\doibase
  10.1103/PhysRevLett.114.016601} {\bibfield  {journal} {\bibinfo  {journal}
  {Phys. Rev. Lett.}\ }\textbf {\bibinfo {volume} {114}},\ \bibinfo {pages}
  {016601} (\bibinfo {year} {2015})}\BibitemShut {NoStop}%
\bibitem [{\citenamefont {Hirsch}(1999)}]{Hirsch1999}%
  \BibitemOpen
  \bibfield  {author} {\bibinfo {author} {\bibfnamefont {J.~E.}\ \bibnamefont
  {Hirsch}},\ }\href {\doibase 10.1103/PhysRevLett.83.1834} {\bibfield
  {journal} {\bibinfo  {journal} {Phys. Rev. Lett.}\ }\textbf {\bibinfo
  {volume} {83}},\ \bibinfo {pages} {1834} (\bibinfo {year}
  {1999})}\BibitemShut {NoStop}%
\bibitem [{\citenamefont {Callen}\ and\ \citenamefont
  {Welton}(1951)}]{Callen1951}%
  \BibitemOpen
  \bibfield  {author} {\bibinfo {author} {\bibfnamefont {H.~B.}\ \bibnamefont
  {Callen}}\ and\ \bibinfo {author} {\bibfnamefont {T.~A.}\ \bibnamefont
  {Welton}},\ }\href {\doibase 10.1103/PhysRev.83.34} {\bibfield  {journal}
  {\bibinfo  {journal} {Phys. Rev.}\ }\textbf {\bibinfo {volume} {83}},\
  \bibinfo {pages} {34} (\bibinfo {year} {1951})}\BibitemShut {NoStop}%
\bibitem [{\citenamefont {Blanter}\ and\ \citenamefont
  {B{\"u}ttiker}(2000)}]{Blanter2000}%
  \BibitemOpen
  \bibfield  {author} {\bibinfo {author} {\bibfnamefont {Y.}~\bibnamefont
  {Blanter}}\ and\ \bibinfo {author} {\bibfnamefont {M.}~\bibnamefont
  {B{\"u}ttiker}},\ }\href {\doibase
  http://dx.doi.org/10.1016/S0370-1573(99)00123-4} {\bibfield  {journal}
  {\bibinfo  {journal} {Physics Reports}\ }\textbf {\bibinfo {volume} {336}},\
  \bibinfo {pages} {1 } (\bibinfo {year} {2000})}\BibitemShut {NoStop}%
\bibitem [{\citenamefont {Beenakker}\ and\ \citenamefont
  {Sch{\"o}nenberger}(2003)}]{Beenakker2003}%
  \BibitemOpen
  \bibfield  {author} {\bibinfo {author} {\bibfnamefont {C.}~\bibnamefont
  {Beenakker}}\ and\ \bibinfo {author} {\bibfnamefont {C.}~\bibnamefont
  {Sch{\"o}nenberger}},\ }\href {\doibase 10.1063/1.1583532} {\bibfield
  {journal} {\bibinfo  {journal} {Physics Today}\ }\textbf {\bibinfo {volume}
  {56}},\ \bibinfo {pages} {37} (\bibinfo {year} {2003})}\BibitemShut {NoStop}%
\bibitem [{\citenamefont {Nazarov}(2003)}]{Nazarov2003}%
  \BibitemOpen
  \bibfield  {author} {\bibinfo {author} {\bibfnamefont {Y.}~\bibnamefont
  {Nazarov}},\ }\href {https://books.google.de/books?id=55Ww6x-SaNcC} {\emph
  {\bibinfo {title} {Quantum Noise in Mesoscopic Physics}}}\ (\bibinfo
  {publisher} {Springer, Netherlands},\ \bibinfo {year} {2003})\BibitemShut
  {NoStop}%
\bibitem [{\citenamefont {Jain}(1989)}]{Jain1989}%
  \BibitemOpen
  \bibfield  {author} {\bibinfo {author} {\bibfnamefont {J.~K.}\ \bibnamefont
  {Jain}},\ }\href {\doibase 10.1103/PhysRevLett.63.199} {\bibfield  {journal}
  {\bibinfo  {journal} {Phys. Rev. Lett.}\ }\textbf {\bibinfo {volume} {63}},\
  \bibinfo {pages} {199} (\bibinfo {year} {1989})}\BibitemShut {NoStop}%
\bibitem [{\citenamefont {Kane}\ and\ \citenamefont {Fisher}(1994)}]{Kane1994}%
  \BibitemOpen
  \bibfield  {author} {\bibinfo {author} {\bibfnamefont {C.~L.}\ \bibnamefont
  {Kane}}\ and\ \bibinfo {author} {\bibfnamefont {M.~P.~A.}\ \bibnamefont
  {Fisher}},\ }\href {\doibase 10.1103/PhysRevLett.72.724} {\bibfield
  {journal} {\bibinfo  {journal} {Phys. Rev. Lett.}\ }\textbf {\bibinfo
  {volume} {72}},\ \bibinfo {pages} {724} (\bibinfo {year} {1994})}\BibitemShut
  {NoStop}%
\bibitem [{\citenamefont {Reznikov}\ \emph {et~al.}(1999)\citenamefont
  {Reznikov}, \citenamefont {Picciotto}, \citenamefont {Griffiths},
  \citenamefont {Heiblum},\ and\ \citenamefont {Umansky}}]{Reznikov1999}%
  \BibitemOpen
  \bibfield  {author} {\bibinfo {author} {\bibfnamefont {M.}~\bibnamefont
  {Reznikov}}, \bibinfo {author} {\bibfnamefont {R.~d.}\ \bibnamefont
  {Picciotto}}, \bibinfo {author} {\bibfnamefont {T.~G.}\ \bibnamefont
  {Griffiths}}, \bibinfo {author} {\bibfnamefont {M.}~\bibnamefont {Heiblum}},
  \ and\ \bibinfo {author} {\bibfnamefont {V.}~\bibnamefont {Umansky}},\ }\href
  {\doibase 10.1038/20384} {\bibfield  {journal} {\bibinfo  {journal} {Nature}\
  }\textbf {\bibinfo {volume} {399}},\ \bibinfo {pages} {238} (\bibinfo {year}
  {1999})}\BibitemShut {NoStop}%
\bibitem [{\citenamefont {Jehl}\ \emph {et~al.}(2000)\citenamefont {Jehl},
  \citenamefont {Sanquer}, \citenamefont {Calemczuk},\ and\ \citenamefont
  {Mailly}}]{Jehl2000}%
  \BibitemOpen
  \bibfield  {author} {\bibinfo {author} {\bibfnamefont {X.}~\bibnamefont
  {Jehl}}, \bibinfo {author} {\bibfnamefont {M.}~\bibnamefont {Sanquer}},
  \bibinfo {author} {\bibfnamefont {R.}~\bibnamefont {Calemczuk}}, \ and\
  \bibinfo {author} {\bibfnamefont {D.}~\bibnamefont {Mailly}},\ }\href
  {\doibase 10.1038/35011012} {\bibfield  {journal} {\bibinfo  {journal}
  {Nature}\ }\textbf {\bibinfo {volume} {405}},\ \bibinfo {pages} {50}
  (\bibinfo {year} {2000})}\BibitemShut {NoStop}%
\bibitem [{\citenamefont {Kamra}\ and\ \citenamefont
  {Belzig}(2016)}]{Kamra2016}%
  \BibitemOpen
  \bibfield  {author} {\bibinfo {author} {\bibfnamefont {A.}~\bibnamefont
  {Kamra}}\ and\ \bibinfo {author} {\bibfnamefont {W.}~\bibnamefont {Belzig}},\
  }\href {\doibase 10.1103/PhysRevLett.116.146601} {\bibfield  {journal}
  {\bibinfo  {journal} {Phys. Rev. Lett.}\ }\textbf {\bibinfo {volume} {116}},\
  \bibinfo {pages} {146601} (\bibinfo {year} {2016})}\BibitemShut {NoStop}%
\bibitem [{\citenamefont {Kittel}(1963)}]{Kittel1963}%
  \BibitemOpen
  \bibfield  {author} {\bibinfo {author} {\bibfnamefont {C.}~\bibnamefont
  {Kittel}},\ }\href {https://books.google.de/books?id=zgBRAAAAMAAJ} {\emph
  {\bibinfo {title} {Quantum theory of solids}}}\ (\bibinfo  {publisher}
  {Wiley, New York},\ \bibinfo {year} {1963})\BibitemShut {NoStop}%
\bibitem [{\citenamefont {Akhiezer}\ \emph {et~al.}(1968)\citenamefont
  {Akhiezer}, \citenamefont {Bar'iakhtar},\ and\ \citenamefont
  {Peletminski}}]{Akhiezer1968}%
  \BibitemOpen
  \bibfield  {author} {\bibinfo {author} {\bibfnamefont {A.}~\bibnamefont
  {Akhiezer}}, \bibinfo {author} {\bibfnamefont {V.}~\bibnamefont
  {Bar'iakhtar}}, \ and\ \bibinfo {author} {\bibfnamefont {S.}~\bibnamefont
  {Peletminski}},\ }\href {http://books.google.nl/books?id=GpA6AAAAMAAJ} {\emph
  {\bibinfo {title} {Spin waves}}}\ (\bibinfo  {publisher} {North-Holland
  Publishing Company, Amsterdam},\ \bibinfo {year} {1968})\BibitemShut
  {NoStop}%
\bibitem [{\citenamefont {Kittel}(1949)}]{Kittel1949}%
  \BibitemOpen
  \bibfield  {author} {\bibinfo {author} {\bibfnamefont {C.}~\bibnamefont
  {Kittel}},\ }\href {\doibase 10.1103/RevModPhys.21.541} {\bibfield  {journal}
  {\bibinfo  {journal} {Rev. Mod. Phys.}\ }\textbf {\bibinfo {volume} {21}},\
  \bibinfo {pages} {541} (\bibinfo {year} {1949})}\BibitemShut {NoStop}%
\bibitem [{\citenamefont {Kamra}\ \emph {et~al.}(2015)\citenamefont {Kamra},
  \citenamefont {Keshtgar}, \citenamefont {Yan},\ and\ \citenamefont
  {Bauer}}]{Kamra2015}%
  \BibitemOpen
  \bibfield  {author} {\bibinfo {author} {\bibfnamefont {A.}~\bibnamefont
  {Kamra}}, \bibinfo {author} {\bibfnamefont {H.}~\bibnamefont {Keshtgar}},
  \bibinfo {author} {\bibfnamefont {P.}~\bibnamefont {Yan}}, \ and\ \bibinfo
  {author} {\bibfnamefont {G.~E.~W.}\ \bibnamefont {Bauer}},\ }\href {\doibase
  10.1103/PhysRevB.91.104409} {\bibfield  {journal} {\bibinfo  {journal} {Phys.
  Rev. B}\ }\textbf {\bibinfo {volume} {91}},\ \bibinfo {pages} {104409}
  (\bibinfo {year} {2015})}\BibitemShut {NoStop}%
\bibitem [{Note1()}]{Note1}%
  \BibitemOpen
  \bibinfo {note} {Strictly speaking, the magnetostatic approximation is not
  valid for a certain narrow range of low k excitations~\cite {Akhiezer1968}.
  However, as we see in the final results, the thermal noise has contribution
  from excitations in a wide k range making the error due to an imprecise
  treatment of a fraction of this range negligible.}\BibitemShut {Stop}%
\bibitem [{Note2()}]{Note2}%
  \BibitemOpen
  \bibinfo {note} {The $\gamma / |\gamma |$ factor, which is often omitted
  assuming positive $\gamma $, is essential for a valid transformation
  consistent with angular momentum conservation.}\BibitemShut {Stop}%
\bibitem [{\citenamefont {Holstein}\ and\ \citenamefont
  {Primakoff}(1940)}]{Holstein1940}%
  \BibitemOpen
  \bibfield  {author} {\bibinfo {author} {\bibfnamefont {T.}~\bibnamefont
  {Holstein}}\ and\ \bibinfo {author} {\bibfnamefont {H.}~\bibnamefont
  {Primakoff}},\ }\href {\doibase 10.1103/PhysRev.58.1098} {\bibfield
  {journal} {\bibinfo  {journal} {Phys. Rev.}\ }\textbf {\bibinfo {volume}
  {58}},\ \bibinfo {pages} {1098} (\bibinfo {year} {1940})}\BibitemShut
  {NoStop}%
\bibitem [{\citenamefont {Walls}\ and\ \citenamefont
  {Milburn}(2008)}]{Walls2008}%
  \BibitemOpen
  \bibfield  {author} {\bibinfo {author} {\bibfnamefont {D.}~\bibnamefont
  {Walls}}\ and\ \bibinfo {author} {\bibfnamefont {G.}~\bibnamefont
  {Milburn}},\ }\href {https://books.google.de/books?id=LiWsc3Nlf0kC} {\emph
  {\bibinfo {title} {Quantum Optics}}}\ (\bibinfo  {publisher} {Springer,
  Berlin},\ \bibinfo {year} {2008})\BibitemShut {NoStop}%
\bibitem [{\citenamefont {Zhang}\ and\ \citenamefont
  {Zhang}(2012)}]{Zhang2012}%
  \BibitemOpen
  \bibfield  {author} {\bibinfo {author} {\bibfnamefont {S.~S.-L.}\
  \bibnamefont {Zhang}}\ and\ \bibinfo {author} {\bibfnamefont
  {S.}~\bibnamefont {Zhang}},\ }\href {\doibase 10.1103/PhysRevB.86.214424}
  {\bibfield  {journal} {\bibinfo  {journal} {Phys. Rev. B}\ }\textbf {\bibinfo
  {volume} {86}},\ \bibinfo {pages} {214424} (\bibinfo {year}
  {2012})}\BibitemShut {NoStop}%
\bibitem [{\citenamefont {Gardiner}\ and\ \citenamefont
  {Collett}(1985)}]{Gardiner1985}%
  \BibitemOpen
  \bibfield  {author} {\bibinfo {author} {\bibfnamefont {C.~W.}\ \bibnamefont
  {Gardiner}}\ and\ \bibinfo {author} {\bibfnamefont {M.~J.}\ \bibnamefont
  {Collett}},\ }\href {\doibase 10.1103/PhysRevA.31.3761} {\bibfield  {journal}
  {\bibinfo  {journal} {Phys. Rev. A}\ }\textbf {\bibinfo {volume} {31}},\
  \bibinfo {pages} {3761} (\bibinfo {year} {1985})}\BibitemShut {NoStop}%
\bibitem [{\citenamefont {Gilbert}(2004)}]{Gilbert2004}%
  \BibitemOpen
  \bibfield  {author} {\bibinfo {author} {\bibfnamefont {T.}~\bibnamefont
  {Gilbert}},\ }\href {\doibase 10.1109/TMAG.2004.836740} {\bibfield  {journal}
  {\bibinfo  {journal} {Magnetics, IEEE Transactions on}\ }\textbf {\bibinfo
  {volume} {40}},\ \bibinfo {pages} {3443} (\bibinfo {year}
  {2004})}\BibitemShut {NoStop}%
\bibitem [{Note3()}]{Note3}%
  \BibitemOpen
  \bibinfo {note} {The ``single-sided'' (also known as ``one-sided'' or
  ``unilateral'') power spectral density $S(\Omega )$ is defined as twice the
  usual power for positive frequencies ($\Omega > 0$) and zero for negative
  frequencies ($\Omega < 0$). The definition is a matter of convenience so that
  in evaluating the total power in a signal via the Parseval theorem~\cite
  {Howard2004}, one needs to integrate over positive frequencies
  only.}\BibitemShut {Stop}%
\bibitem [{\citenamefont {Howard}(2004)}]{Howard2004}%
  \BibitemOpen
  \bibfield  {author} {\bibinfo {author} {\bibfnamefont {R.}~\bibnamefont
  {Howard}},\ }\href {https://books.google.de/books?id=gOqmuPP7dqgC} {\emph
  {\bibinfo {title} {Principles of Random Signal Analysis and Low Noise Design:
  The Power Spectral Density and its Applications}}}\ (\bibinfo  {publisher}
  {Wiley-Interscience, New York},\ \bibinfo {year} {2004})\BibitemShut
  {NoStop}%
\bibitem [{\citenamefont {Czeschka}\ \emph {et~al.}(2011)\citenamefont
  {Czeschka}, \citenamefont {Dreher}, \citenamefont {Brandt}, \citenamefont
  {Weiler}, \citenamefont {Althammer}, \citenamefont {Imort}, \citenamefont
  {Reiss}, \citenamefont {Thomas}, \citenamefont {Schoch}, \citenamefont
  {Limmer}, \citenamefont {Huebl}, \citenamefont {Gross},\ and\ \citenamefont
  {Goennenwein}}]{Czeschka2011}%
  \BibitemOpen
  \bibfield  {author} {\bibinfo {author} {\bibfnamefont {F.~D.}\ \bibnamefont
  {Czeschka}}, \bibinfo {author} {\bibfnamefont {L.}~\bibnamefont {Dreher}},
  \bibinfo {author} {\bibfnamefont {M.~S.}\ \bibnamefont {Brandt}}, \bibinfo
  {author} {\bibfnamefont {M.}~\bibnamefont {Weiler}}, \bibinfo {author}
  {\bibfnamefont {M.}~\bibnamefont {Althammer}}, \bibinfo {author}
  {\bibfnamefont {I.-M.}\ \bibnamefont {Imort}}, \bibinfo {author}
  {\bibfnamefont {G.}~\bibnamefont {Reiss}}, \bibinfo {author} {\bibfnamefont
  {A.}~\bibnamefont {Thomas}}, \bibinfo {author} {\bibfnamefont
  {W.}~\bibnamefont {Schoch}}, \bibinfo {author} {\bibfnamefont
  {W.}~\bibnamefont {Limmer}}, \bibinfo {author} {\bibfnamefont
  {H.}~\bibnamefont {Huebl}}, \bibinfo {author} {\bibfnamefont
  {R.}~\bibnamefont {Gross}}, \ and\ \bibinfo {author} {\bibfnamefont
  {S.~T.~B.}\ \bibnamefont {Goennenwein}},\ }\href {\doibase
  10.1103/PhysRevLett.107.046601} {\bibfield  {journal} {\bibinfo  {journal}
  {Phys. Rev. Lett.}\ }\textbf {\bibinfo {volume} {107}},\ \bibinfo {pages}
  {046601} (\bibinfo {year} {2011})}\BibitemShut {NoStop}%
\bibitem [{\citenamefont {Boona}\ and\ \citenamefont
  {Heremans}(2014)}]{Boona2014}%
  \BibitemOpen
  \bibfield  {author} {\bibinfo {author} {\bibfnamefont {S.~R.}\ \bibnamefont
  {Boona}}\ and\ \bibinfo {author} {\bibfnamefont {J.~P.}\ \bibnamefont
  {Heremans}},\ }\href {\doibase 10.1103/PhysRevB.90.064421} {\bibfield
  {journal} {\bibinfo  {journal} {Phys. Rev. B}\ }\textbf {\bibinfo {volume}
  {90}},\ \bibinfo {pages} {064421} (\bibinfo {year} {2014})}\BibitemShut
  {NoStop}%
\bibitem [{\citenamefont {Kikkawa}\ \emph {et~al.}(2015)\citenamefont
  {Kikkawa}, \citenamefont {Uchida}, \citenamefont {Daimon}, \citenamefont
  {Qiu}, \citenamefont {Shiomi},\ and\ \citenamefont {Saitoh}}]{Kikkawa2015}%
  \BibitemOpen
  \bibfield  {author} {\bibinfo {author} {\bibfnamefont {T.}~\bibnamefont
  {Kikkawa}}, \bibinfo {author} {\bibfnamefont {K.-i.}\ \bibnamefont {Uchida}},
  \bibinfo {author} {\bibfnamefont {S.}~\bibnamefont {Daimon}}, \bibinfo
  {author} {\bibfnamefont {Z.}~\bibnamefont {Qiu}}, \bibinfo {author}
  {\bibfnamefont {Y.}~\bibnamefont {Shiomi}}, \ and\ \bibinfo {author}
  {\bibfnamefont {E.}~\bibnamefont {Saitoh}},\ }\href {\doibase
  10.1103/PhysRevB.92.064413} {\bibfield  {journal} {\bibinfo  {journal} {Phys.
  Rev. B}\ }\textbf {\bibinfo {volume} {92}},\ \bibinfo {pages} {064413}
  (\bibinfo {year} {2015})}\BibitemShut {NoStop}%
\bibitem [{\citenamefont {Ritzmann}\ \emph {et~al.}(2015)\citenamefont
  {Ritzmann}, \citenamefont {Hinzke}, \citenamefont {Kehlberger}, \citenamefont
  {Guo}, \citenamefont {Kl\"aui},\ and\ \citenamefont {Nowak}}]{Ritzmann2015}%
  \BibitemOpen
  \bibfield  {author} {\bibinfo {author} {\bibfnamefont {U.}~\bibnamefont
  {Ritzmann}}, \bibinfo {author} {\bibfnamefont {D.}~\bibnamefont {Hinzke}},
  \bibinfo {author} {\bibfnamefont {A.}~\bibnamefont {Kehlberger}}, \bibinfo
  {author} {\bibfnamefont {E.-J.}\ \bibnamefont {Guo}}, \bibinfo {author}
  {\bibfnamefont {M.}~\bibnamefont {Kl\"aui}}, \ and\ \bibinfo {author}
  {\bibfnamefont {U.}~\bibnamefont {Nowak}},\ }\href {\doibase
  10.1103/PhysRevB.92.174411} {\bibfield  {journal} {\bibinfo  {journal} {Phys.
  Rev. B}\ }\textbf {\bibinfo {volume} {92}},\ \bibinfo {pages} {174411}
  (\bibinfo {year} {2015})}\BibitemShut {NoStop}%
\bibitem [{\citenamefont {Sharma}\ \emph {et~al.}()\citenamefont {Sharma},
  \citenamefont {Kamra}, \citenamefont {Xiao},\ and\ \citenamefont
  {Bauer}}]{Sharma2016}%
  \BibitemOpen
  \bibfield  {author} {\bibinfo {author} {\bibfnamefont {S.}~\bibnamefont
  {Sharma}}, \bibinfo {author} {\bibfnamefont {A.}~\bibnamefont {Kamra}},
  \bibinfo {author} {\bibfnamefont {J.}~\bibnamefont {Xiao}}, \ and\ \bibinfo
  {author} {\bibfnamefont {G.~E.~W.}\ \bibnamefont {Bauer}},\ }\href@noop {}
  {\bibinfo  {journal} {unpublished}\ }\BibitemShut {NoStop}%
\end{thebibliography}%

\end{document}